\documentclass[aps,prd,eqsecnum,superscriptaddress,twocolumn,nofootinbib,floatfix,preprintnumbers]{revtex4-2}

\usepackage{url,hyperref}
\usepackage{ulem}
\usepackage{aas_macros}
\usepackage{multirow}
\usepackage{graphicx}
\usepackage{footnote}
\usepackage{natbib}
\usepackage[fleqn]{amsmath}
\usepackage{amsmath}
\usepackage{color}
\usepackage[title,toc,page]{appendix}
\usepackage{amssymb}
\usepackage{xcolor}
\usepackage{subcaption}
\usepackage{hyphenat}
\usepackage{ORCIDinREVTeX}
\usepackage{placeins}
\usepackage{tablefootnote}
\usepackage{comment}

\begin{document}

\title{Long-duration GW Searches for Sub-solar NSs and Superkilonovae using CoCoA}

\newcommand*{\JHU}{William H. Miller III Department of Physics and Astronomy, Johns Hopkins University, Baltimore, MD 21218, USA}\affiliation{\JHU}
\newcommand*{\Cardiff}{Gravity Exploration Institute, School of Physics and Astronomy, Cardiff University, Cardiff, CF24 3AA, United Kingdom }\affiliation{\Cardiff}
\newcommand*{\ARCO}{Astrophysics Research Center of the Open University (ARCO), Department of Natural Sciences, Ra’anana 4353701, Israel}\affiliation{\ARCO}
\newcommand*{\URI}{Department of Physics, University of Rhode Island, Kingston, Rhode Island 02881 (USA)}\affiliation{\URI}

\definecolor{darkgreen}{rgb}{0,0.5,0}
\newcommand{\divya}[1]{\small{\textcolor{darkgreen}{\sf{[DJ: #1]}}}}
\newcommand{\djtext}[1]{\textcolor{darkgreen}{#1}}

\author{Abby Tejera}\orcid{0009-0009-1694-5328}\affiliation{\JHU}
\author{Divyajyoti}\orcid{0000-0000-0000-0000}\affiliation{\Cardiff}
\author{Alessandra Corsi}\orcid{0000-0000-0000-0000}\affiliation{\JHU}
\author{Yossef Zenati}\orcid{0000-0002-0632-8897}\affiliation{\ARCO} \affiliation{\JHU}
\author{Robert Coyne}\orcid{0000-0000-0000-0000}\affiliation{\URI}
\author{Logan Cote}\orcid{0000-0000-0000-0000}\affiliation{\URI}

\date[\relax]{compiled \today }

\begin{abstract}
On 2025 August 18, the LIGO–Virgo–KAGRA collaboration reported a sub-threshold gravitational-wave (GW) candidate, S250818k, consistent with a binary neutron star (NS) merger potentially involving a sub-solar-mass compact object. 
Follow-up electromagnetic (EM) observations identified a Type IIb supernova, SN~2025ulz within the broad localization area of the GW signal. This potential link between a sub-solar GW event candidate and a Type IIb SN, while not confirmed given the low statistical significance of S250818k, has nonetheless sparked renewed interest in the ``superkilonova'' scenario, where sub-solar-mass NSs form through processes like the fragmentation of an accretion disk or core fission in a collapsing star. In this picture, the in-spiral and merger of a sub-solar NS–NS binary is followed by the merger of the NS-NS remnant with the central black hole (BH), producing chirp-like GW signals. For sub-solar NSs with masses in the $(0.1$–$1)\,{\rm M}_{\odot}$ range and BH masses in the so-called lower mass gap range of $\approx (3$–$5)\,{\rm M}_{\odot}$, these signals can persist in the 20 - 1024\,Hz frequency band of ground-based GW detectors for ${\cal O}(10^2$–$10^3)\,{\rm s}$, potentially offering an opportunity to probe the superkilonova scenario, as well as the lower mass gap between NSs and stellar-mass BHs. However, the complexity of the underlying astrophysics may yield waveforms that deviate from standard templates, limiting the use of matched filtering in real GW searches.
We therefore explore the detectability of such signals using the Cross-Correlation Algorithm (\texttt{CoCoA}), a more robust though less sensitive cross-correlation method. We discuss general strategies for implementing \texttt{CoCoA} superkilonova searches via either targeted follow-up of candidate chirps identified in matched-filter searches, or EM-triggered searches of stripped-envelope core-collapse SNe.  
\end{abstract}

\flushbottom 
\maketitle

\thispagestyle{empty}

\section{Introduction} \label{sec:intro}
The fourth observing run (O4) of the LIGO \citep{2015CQGra..32g4001L}, Virgo \citep{2015CQGra..32b4001A}, and KAGRA \citep{2021PTEP.2021eA101A} detectors has provided new insights into the diversity of compact binary mergers \citep{2025arXiv250818083T,LIGO+25_Rates}. While O4 did not yield a GW170817-like \citep{2017PhRvL.119p1101A,2017ApJ...848L..13A,2017ApJ...848L..12A} high-significance binary neutron star (NS) merger with a confirmed electromagnetic (EM) counterpart, it enabled the identification of a particularly intriguing event.
In August 2025, the sub-threshold compact binary candidate S250818k was reported in data from the LIGO detectors \citep{2025GCN.41437....1L,GCN41440}. Low-latency parameter estimation placed this candidate in the sub-solar mass regime, with a chirp mass of $\simeq 0.87\,M_\odot$, implying at least one component with a mass below $1\,M_\odot$ \citep{KasliwalM+25_S250818k,HallX+25_AT2025ulz_S250818k}. Extensive EM follow-up observations covered a large fraction of the localization region and led to the identification by the Zwicky Transient Facility (ZTF) of ZTF25abjmnps \citep{2025GCN.41414....1S} as the only candidate exhibiting both a red color (reminiscent of a kilonova-like emission) and a photometric redshift consistent with the GW distance, motivating expectations of a possible gamma-ray burst (GRB) association. Subsequent observations established the transient as a stripped-envelope Type IIb supernova (SN) at $z=0.0848$ \citep{PassalevaN+25_NIR, AnJ+25, AnguloC+25_AT2025ulz, AntierS+25, BanerjeeS+25_SNeII,  GillandersH+25_PanStarrs, KasliwalM+25_KeckI, YangYu+26_25ulz}.
While early-time X-ray and radio observations did not yield detections \citep{OConnorB+25_25ulz, HallJ+25_Swift,2025ApJ...994L..45F}, late-time radio observations showed emission consistent with a Type IIb SN, though the presence of a faint off-axis GRB jet remains compatible with the data \citep{2026arXiv260405128O}.

While the association between S250818k and SN2025ulz remains uncertain and highly debated, it may nonetheless represent a qualitatively novel target motivating further studies in both the observational and theoretical domains. Indeed, given the potential link between S250818k and a sub-solar-mass NS event, several authors have interpreted this system within a ``superkilonova'' scenario, in which sub-solar NSs form through core fission or accretion-disk fragmentation within a core-collapse SN (CCSN). More broadly, the superkilonova class has been proposed to include CCSNe exhibiting embedded kilonova-like $r$-process nucleosynthesis signatures \citep{Piro_Pfhal07,Metzger_Hui_Cantiello24,Chen_Metzger25}. 

Recent theoretical work offers important insight into how sub-solar-mass compact objects may form in core collapses. Zenati et al. 2025 \cite{Zenati+25} demonstrated that mass transfer in eccentric BH–NS mergers can alter orbital evolution and significantly reduce the NS mass when the transfer rate exceeds $\geq 10^{-2}{\rm M}_{\odot}$ per orbit, forming an accretion disk around the BH before merger with mass similar to post-merger debris ($\sim 0.1 {\rm M}_{\odot}$). This mechanism enables the creation of subsolar-mass compact objects. Complementing this, Wu et al. 2026 ~\cite{WuJ+26} examined the GW signatures of hierarchical subsolar-mass mergers resulting from fragmentation within collapsar accretion disks. Through numerical relativity simulations, it was demonstrated that repeated capture and the merger dynamics can impart substantial orbital eccentricity $(e_0 \simeq 0.6$) to the final merger. Overall. these and other studies \cite{Piro_Pfhal07,Metzger_Hui_Cantiello24,Chen_Metzger25} offer a theoretical base supporting the superkilonova scenario and highlight the need for specialized GW search strategies.

In the superkilonova framework, the in-spiral and merger of a sub-solar NS-NS binary formed within the SN disk would be followed by the in-spiral and merger of the remnant with the central black hole (for alternative formation channels and BH–NS scenarios, see \citep{BelczynskiK+02,DaviesM+05,Piro_Pfhal07,BelczynskiK+16,Metzger_Hui_Cantiello24,Zenati+25,StegmannJ_KlenckiJ25}), producing long-duration, chirp-like GW signals. For NS masses in the $(0.1$–$1)\,{\rm M}_{\odot}$ range and BH masses in the $(3$–$5)\,{\rm M}_{\odot}$ range, these signals can persist in the most sensitive frequency range of ground-based GW detectors for ${\cal O}(10^2$–$10^3\,{\rm s})$, thereby also probing the lower mass gap between NSs and stellar-mass BHs.
While ground-based GW detectors may be sensitive to such superkilonova signals, existing detectability estimates typically assume optimal matched-filtering techniques \citep{Chen_Metzger25}. However, the complexity of the underlying astrophysics may produce waveforms that deviate significantly from standard templates, thereby reducing the effectiveness of matched-filter searches and/or potentially producing sub-threshold triggers \cite{2026arXiv260505444T,Cornish:2026ltu}.

Motivated by the above considerations, here we estimate the detectability of GW signals from compact binary coalescences involving sub-solar objects using the Cross-Correlation Algorithm (\texttt{CoCoA}) \cite{CoyneR+2016}, a more robust method optimized for long-duration GW signals such as those predicted in this scenario. 
Our paper is organized as follows. In Section~\ref{sec:rates} we discuss potential bounds on the rates of superkilonovae, which directly impact what can be defined as an astrophysically interesting distance horizon for any GW search. In Section~\ref{sec:gw-waveforms} we describe the approximate waveforms that we use to assess the GW detectability. In Section \ref{sec:methods} we describe the methods we use to assess GW detectability in stochastic searches, addressing both targeted follow-up of low-significance (potentially affected by waveform mismatch) candidate chirps identified in matched-filter searches, and EM-triggered searches of stripped-envelope CCSNe \citep{KasliwalM+25_S250818k,HallX+25_AT2025ulz_S250818k}. In Section \ref{sec:det} we discuss the detector networks and sensitivity curves used in this study, followed by Section \ref{sec:results} where we present our results. Finally, in Section~\ref{sec:conclusion}, summarize and conclude.

\section{Rates} \label{sec:rates}
The superkilonova scenario posits that sub-solar-mass NSs can form within collapsar accretion disks via fragmentation or core fission, and subsequently merge with each other and then with the central BH \citep{Piro_Pfhal07,Metzger_Hui_Cantiello24, Chen_Metzger25}. Furthermore, Wu et al. (2026) \cite{WuJ+26} find that fragmentation within an accretion disk can generate smaller NS fragments capable of hierarchical merging, which may result in observable eccentricities in their orbital trajectories. The rate of such events is highly uncertain, as it depends on the fraction of collapsars with sufficient disk mass and angular momentum to fragment, on the probability that fragmentation produces bound sub-solar NS binaries, which in turn is impacted by the timescale for in-disk merger relative to disk dispersal. Hereafter, we establish a reasonable upper bound on the intrinsic rate of superkilonovae one might expect in the local universe, so as to understand what horizon distances may be relevant for potentially probing these events via GW observations. 

First, in a superkilonova scenario, the collapsar disk must become gravitationally unstable and satisfy the cooling conditions required for fragmentation; we denote the fraction of collapsars satisfying these conditions by ($f_{frag}$). Among the fragmenting disks, only a fraction ($f_{\geq 2{\rm NS}\mid{\rm frag}}$) may produce at least two surviving subsolar-mass NSs. These NSs must subsequently form a bound binary, a condition that we described by ($f_{\rm pair\mid\geq 2{\rm NS}}$). This pair then needs to merge before the cooling disk or infall into the central BH occur ($f_{\rm merge\mid pair}$). Putting it all together we can write:
\begin{equation}
\begin{aligned}
f_{\rm SKN|coll}
={}&
f_{\rm frag}\times
f_{\geq 2{\rm NS}\mid{\rm frag}}\times
f_{\rm pair\mid\geq 2{\rm NS}}\times
f_{\rm merge\mid pair}.
\end{aligned}
\label{eq}
\end{equation}
The branching fraction $f_{\rm SKN|coll}$ is presently unconstrained.
The simulations of \citet{Chen_Metzger25} demonstrate that sufficiently
massive, rapidly cooling collapsar disks can fragment into sub-solar-mass
NSs, while \citet{WuJ+26} show that subsequent hierarchical
mergers can produce distinctive eccentric GW signals. However, neither calculation determines the population-averaged fraction of collapsars that fragment, form at least two surviving NSs, produce a bound binary, and undergo a merger before the disk disperses. This unknown branching fraction therefore dominates the uncertainty in the predicted superkilonova rate, which we can thus write as:
\begin{equation}
\begin{aligned}
R_{0,\rm SKNe}
=f_{\rm SKN\mid coll} \times R_{0,\rm coll}^{\rm true}.
\end{aligned}
\label{eq:rates}
\end{equation}
The tentative association between S250818k and AT2025ulz does not presently provide a calibration of this branching fraction \citep{KasliwalM+25_KeckI,KasliwalM+25_S250818k,HallX+25_AT2025ulz_S250818k}. Hence, optimistically, hereafter 
we  assume that superkilonovae occur in $f_{\rm SKN|coll}=10^{-2}$ as a fiducial benchmark rather than an empirically measured rate.

In the local universe, the comoving event-rate density of CCSNe is well-established at $R_{0,\rm CCSN} \approx 7 \times 10^4$\,Gpc$^{-3}$\,yr$^{-1}$ \citep{RozwadowskaK+21}. This rate provides the baseline from which all compact-object formation channels emerge. Within the CCSN population, the observed on-axis rate of long-duration GRBs (LGRBs) associated with collapsars is $R^{\rm obs}_{0,\rm LGRB} \simeq 1.3^{+0.6}_{-0.7}$ Gpc$^{-3}$ yr$^{-1}$ \citep{Wanderman_Piran10,SunH+15,LloydRonningN+19}. Low-luminosity GRBs (LL-GRBs), which may represent a distinct collapsar mode or viewing-angle effects, occur at higher rates of $R^{\rm obs}_{0,\rm LLGRB} \sim$ few $\times 10^2$\,Gpc$^{-3}$\,yr$^{-1}$ \citep{LiangE+07,PescalliA+16,RuffiniR+18,DongX+23,CamisascaA+23}. However, their intrinsic rate is bounded by the rate of broad-lined Type Ic SNe, which are estimated to represent $\approx 4.4\%$ of CCSN with rates of $\sim 3\times 10^4$\,Gpc$^{-3}$ yr$^{-1}$ \citep{JapeljJ+18,PessiT+25}. Recent radio studies of broad-lined Type Ic SNe have searched for relativistic ejecta indicative of engine-powered explosions \citep{MarguttiR+14,2016ApJ...830...42C,CorsiA+23,ODwyerT+26}. These observations have constrained the fraction of engine-powered events similar to the broad-lined Type Ic SN\,1998bw, associated with the LL-GRB\,980425, to $<16\%$ \citep{ODwyerT+26}, implying that successful LL collapsar jets constitute $\lesssim 1$\% of all core collapses \citep{2016ApJ...830...42C,CorsiA+23,ODwyerT+26}, or $R^{\rm true}_{0,\rm LLGRB}\lesssim 7
\times 10^2$\,Gpc$^{-3}$\,yr$^{-1}$.
This, in turn, implies that the true rate $R^{\rm true}_{0,\rm LLGRB}$ cannot be much larger than $R^{\rm obs}_{0,\rm LLGRB}$. 

The above constraints, combined with the assumption that at most 1\% of collapsars can power superkilonova (or $f_{\rm SKN|coll}\sim 10^{-2}$, see Equation \ref{eq:rates}), imply:
 \begin{equation}
R_{0,\rm SKNe} \lesssim 7\,\text{Gpc}^{-3}\,\text{yr}^{-1}.
\label{eq:SKNrate}
\end{equation}
For comparison, the most recent constraint from the LIGO-Virgo-KAGRA's (LVK) O4 run places the local NS-NS merger rate at $R_{0,\rm BNS} = 5.1–154.7$\,Gpc$^{-3}$\,yr$^{-1}$ \citep{LIGO+23_Rates,LIGO+25_Rates,GWTC5_LIGO2026}, and that of subsolar-mass compact object coalescences at $R_{0,\rm Sub-BNS} < 86\ {\rm Gpc^{-3}\,yr^{-1}}$ \cite{Sub_BNSLVK26, PhysRevLett.123.161102, PhysRevLett.129.061104, LVK_subsolar_O3b}.

With the upper bound on the intrinsic rate of superkilonovae provided by Equation \ref{eq:SKNrate}, it is evident that GW searches for potentially associated GW signals need to reach horizon distances of $\gtrsim 470$\,Mpc ($\gtrsim 218$\,Mpc) to ensure a $95\%$ probability of at least one GW detection in a 1-year (10-year)-long GW run.

\section{GW signal modeling} 
\label{sec:gw-waveforms}
To obtain order-of-magnitude estimates of the basic properties of GW signals from sub-solar NS components (their duration in a given frequency range, time-frequency evolution, and amplitude) we use the approximate waveforms described below. Such signals could be targeted by long-duration GW searches like \texttt{CoCoA}. Analytical waveform approximations are computationally cheaper to evaluate than numerical ones, making them well suited for large template banks and for search methods robust enough not to require more accurate numerical waveforms. As we show in what follows, the approximants considered here can be used interchangeably for a \texttt{CoCoA} analysis.

\subsection{Analytical in-spiral waveforms}
\label{sec:analyticalwave}

\begin{figure*}[t]
    \centering
    \includegraphics[width=\linewidth]{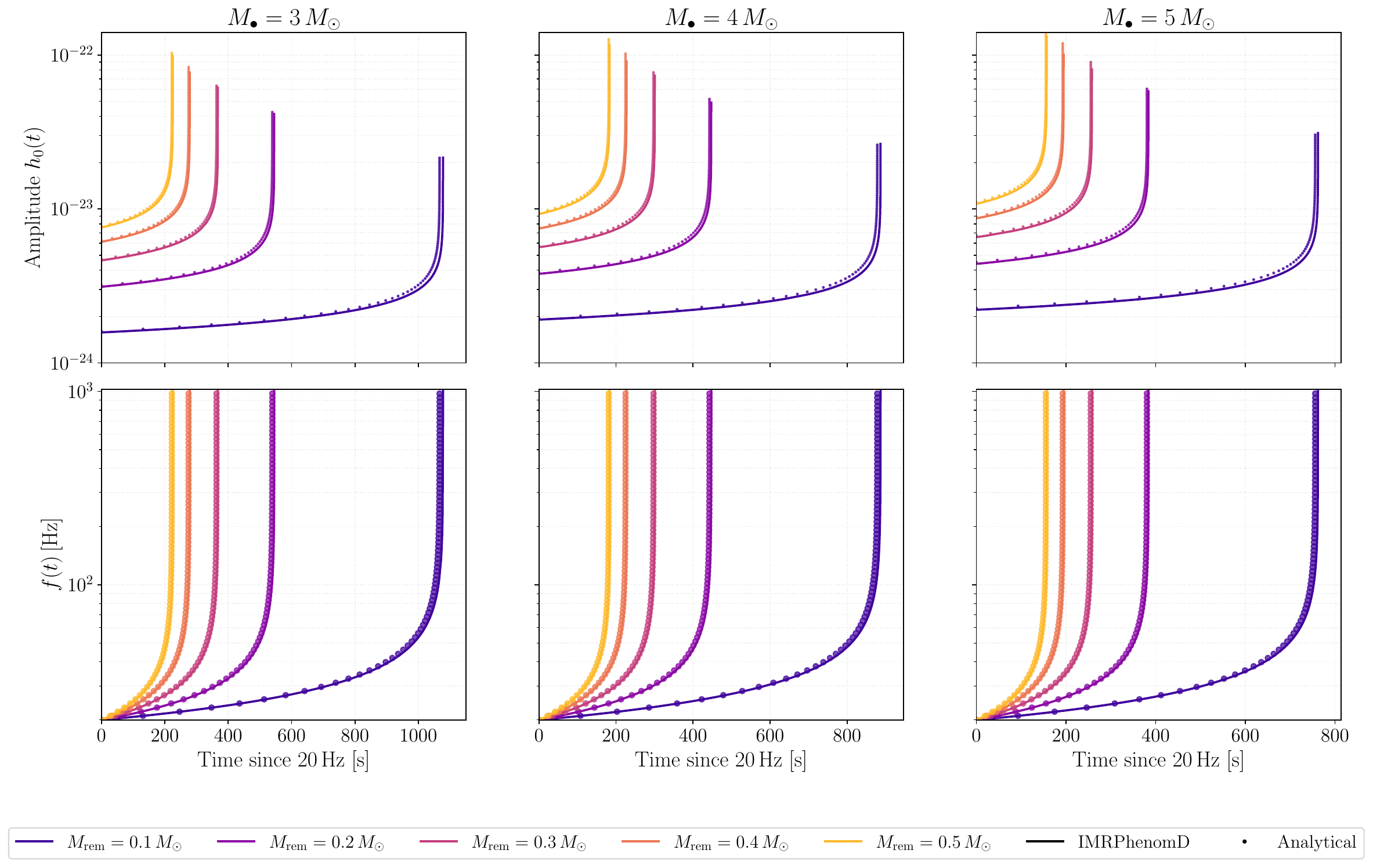}
    \caption{TOP: GW strain amplitude as a function of in-band time (where we take $t=0$ at 20 Hz) for NS-BH binaries at a luminosity distance of $d_L=112$Mpc. BOTTOM: Instantaneous frequency of GW binaries as a function of in-band time. The mass of primary is fixed to (3, 4, 5) $M_\odot$ for left, center, and right columns respectively. The mass of the secondary is taken in the range [0.1, 0.5]~$M_\odot$ and shown with different colours. We have used the LAL waveform \textsc{IMRPhenomD} \citep{lalsuite} to generate the waveforms shown here as solid lines. Analytical waveforms derived from \citet{Chen_Metzger25} (dots) are over-plotted for comparison.}
\label{fig:combined_analytical_LAL_waveforms}
\end{figure*}

We reproduce the approximate analytical waveforms derived in Appendix B of \citet{Chen_Metzger25} for the hierarchical merger system consisting of a sub-solar NS--NS inner binary embedded within a collapsar accretion disk, whose merged remnant subsequently inspirals into the central black hole (BH; outer binary).

For a quasi-circular binary with chirp mass $\mathcal{M} = (m_1 m_2)^{3/5}/(m_1+m_2)^{1/5}$, 
the instantaneous GW frequency for $0\le t < \tau_m$ is: 
\begin{equation}
    f(t) = \frac{1}{8\pi}
            \left(\frac{5}{\tau_m-t}\right)^{3/8}
            \left(\frac{c^3}{G\mathcal{M}}\right)^{5/8},
    \label{eq:freq}
\end{equation} where the GW-driven merger time for a circular binary with semi-major axis $a_0$ is \citep{Peters1964} 
\begin{equation}
    \tau_{m} = \frac{5}{256}\,
                    \frac{c^5\,a_0^4}{G^3\,m_1\,m_2\,(m_1+m_2)}\approx \frac{5}{256}\,
                    \frac{c^5\,a_0^4}{G^3\,m^2_1\,m_2},
    \label{eq:tmerge} 
\end{equation}

 where in the last approximation we have assumed $m_1\gg m_2$ \footnote{This post-Newtonian (PN) result is accurate for the inspiral phase but breaks down in the final stages of merger where full general relativistic effects become important.}.

In the bottom panels of Figure \ref{fig:combined_analytical_LAL_waveforms} we evaluate Eq. \eqref{eq:freq} for the outer NS--BH in-spiral with chirp mass $\mathcal{M}_{\rm out} = \mathcal{M}(M_{\rm rem}, M_\bullet)$, and using Equation \ref{eq:tmerge} with $m_2=M_{\rm rem} = 2M_{\rm NS}$ for the mass of the inner NS--NS merger remnant and $m_1=M_\bullet$ for the BH mass. The BH mass is taken as $M_\bullet \in \{3, 4, 5\} M_\odot$ and the mass of the NS taken in the range $M_\text{NS} \in [0.1, 0.5] M_\odot$. The outer NS–BH binary is taken to originate at a separation $a_{0, out} = r_0 = 200\,R_g$ ($R_g \equiv GM_\bullet/c^2$), motivated by the characteristic outer radius of the gravitationally unstable region in the \citet{Chen_Metzger25} disk simulations. For our mass grid this separation corresponds to a GW frequency of a few Hz, below the 20 Hz low-frequency cutoff of ground-based detectors; we therefore begin tracking and plotting the waveform (Figure \ref{fig:combined_analytical_LAL_waveforms}) from the later point at which the inspiral reaches $f = 20$ Hz.
We note that the maximum possible initial separation for the inner NS--NS binary is set by the Hill radius of the remnant within the disk, 
\begin{equation}
    R_H = r_0\left(\frac{M_{\rm rem}}{3M_\bullet}\right)^{1/3},
    \label{eq:hill} 
\end{equation} 
which is the stability boundary beyond which the binary would be tidally disrupted by the central BH \citep[see Eq. B3 in][]{Chen_Metzger25}.  
Assuming $a_{0, in} = R_H/10$ ensures that the NS--NS merger occurs before the outer inspiral completes, consistent with the hierarchical superkilonova picture \cite{AntogniniJ+14,NaozS2016, 83j3-pgk1, BaiottiL_RezzollaL17,RodriguezL+18}. 

Although before the inner NS-NS merger the total GW strain is the superposition of the two inspiral components, with each component having plus and cross polarization signals of the form: 
\begin{eqnarray}
    h_+(t) \propto  h_0(t)\,
    \cos\!\left(2\pi\int_0^t f(t')\,dt'\right),\\
    h_{\times}(t) \propto  h_0(t)\,
    \sin\!\left(2\pi\int_0^t f(t')\,dt'\right),
\end{eqnarray} 
the GW amplitude $h_0(t)$ of the inner NS-NS merger is expected to be much smaller than that of the BH-NS remnant merger for BH masses in the range 3--5\,M$_{\odot}$ merging with sub-solar NSs (see Eq. B9 in \citep{Chen_Metzger25}). The amplitude (see Eq. B5 in \citep{Chen_Metzger25}):
\begin{equation}
    h_0 (t) = \frac{4\left(G\mathcal{M}\right)^{5/3}}{c^4\,D}
    \bigl(\pi f(t) \bigr)^{2/3},
    \label{eq:strain} 
\end{equation} 
 describes the instantaneous amplitude of the GW signal, and it is plotted in the top panel of Figure \ref{fig:combined_analytical_LAL_waveforms} for a system placed at a distance of $d_L=112$\,Mpc.

 We note that the effects of tidal deformability can significantly modify the waveform, increasing its detectability for a binary with masses $m_1=1.2 M_\odot$ \, and $m_2=1.0 M_\odot$ \citep{Chen_Metzger25}. Hereafter, we neglect tidal deformability, so our detectability estimates can be regarded as conservative. 

\subsection{LAL waveforms}
\label{sec:lalwave}

\begin{figure}
    \centering
    \includegraphics[width=\linewidth]{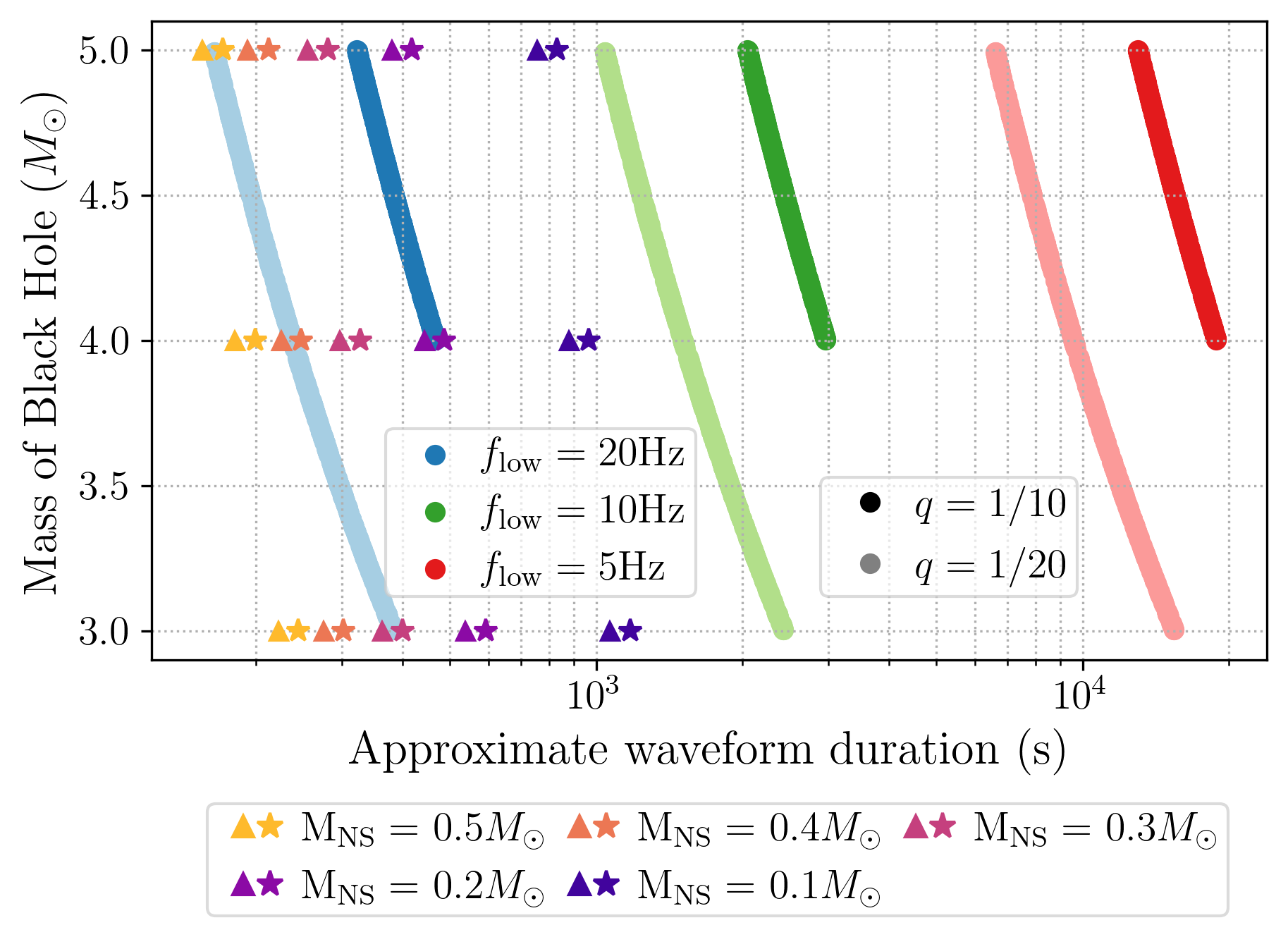}
    \caption{Approximate GW waveform durations for sub-solar NS-BH systems. Stars show the approximate durations (Eq.\,\ref{eq:approxdur}) at $f_{\rm low}$=20\,Hz for the 15 LAL analytical waveforms considered in this study; triangles show the time the corresponding waveforms (Section\,\ref{sec:analyticalwave}) spend between 20 and 1024\,Hz. Tracks in shades of blue, green, and red show approximate LAL durations for a population of NS-BH systems with mass ratios of 1/10 and 1/20, assuming $f_{\rm low}=5, 10$, and $20$\,Hz.}
    \label{fig:wf_len_LAL}
\end{figure}

For direct comparison with the analytical estimates presented in the previous Section, we generate 15 gravitational waveforms for NS-BH binaries with the mass of the BH taken as $M_\bullet \in \{3, 4, 5\} M_\odot$ and the mass of the NS taken in the range $M_\text{NS} \in [0.1, 0.5] M_\odot$, placed at a distance of $d_L=112$~Mpc. We use the LAL waveform model \textsc{IMRPhenomD} \citep{Khan:2015jqa, Husa:2015iqa, lalsuite} to get the time-domain strain. Let $h_+(t)$ and $h_\times(t)$ be the plus and cross polarisations of the GW strain. Then, 
\begin{equation}
    h(t) = h_+(t) - i h_\times(t) = A(t) e^{-i\phi(t)},
\end{equation}
where $A(t)$ is the amplitude of the waveform and $\phi(t)$ denotes the phase. Hence, the instantaneous GW frequency can be written as:
\begin{equation}
    f_\text{inst}^\text{GW}(t) = \frac{1}{2\pi} \frac{\text{d}\phi(t)}{\text{d}t}.
\end{equation}
In Figure~\ref{fig:combined_analytical_LAL_waveforms}, we show the GW amplitude and (instantaneous) frequency evolution, of the NS-BH systems considered here, as a function of time in the top and bottom panels respectively. The mass of the primary is kept constant at $3M_\odot, 4M_\odot$, and $5M_\odot$ for the left, center, and right columns respectively while the mass of the secondary is varied between [0.1, 0.5]~$M_\odot$. We consider $t = 0$ at 20\,Hz which is typically taken as start frequency for observation of signals in the current generation GW detectors. It can be seen in Figure~\ref{fig:combined_analytical_LAL_waveforms} that the lighter systems spend hundreds of seconds in the detector band. We then calculate the approximate waveform duration for these systems using just the 0PN term for \textit{chirp time}~\citep{Allen:2005fk} given by:
\begin{equation}
    T_\text{chirp} = \frac{5}{256\eta} \frac{GM}{c^3} \left(\frac{GM}{c^3}\pi f_\text{low}\right)^{-8/3},
    \label{eq:approxdur}
\end{equation}
where $M$ is the total mass of the binary in solar mass, $\eta= m_1 m_2 / M^2$ is the symmetric mass ratio, and $f_\text{low}$ is the lower cut-off frequency from where the chirp time is calculated. In order to include the possible ringdown, we have multiplied $T_\text{chirp}$ by a fudge factor of 1.1 to get the approximate waveform duration.  In Figure~\ref{fig:wf_len_LAL}  we show with star symbols the approximate waveform durations calculated using Eq. (\ref{eq:approxdur}) and $f_\text{low} = 20$\,Hz. For comparison, we plot with triangles the time the analytical waveforms described in Section \ref{sec:analyticalwave} spend between 20\,Hz and 1024\,Hz.

Next, we simulate a population of NS-BH systems with mass ratios 1/10 and 1/20. For each mass ratio, we sample uniformly in chirp mass such that the primary mass lies between $m_1 \in [3, 5]~M_\odot$ and the secondary mass lies between $m_2 \in [0.2, 1]~M_\odot$. In Figure~\ref{fig:wf_len_LAL} we show the approximate waveform durations for this LAL population (shades of blue, green, and red tracks) when assuming $f_\text{low} = 5, 10, 20$\,Hz so as to show the impact of the increased low-frequency sensitivity on long-duration GW searches with next-generation ground-based GW detectors (Section \ref{sec:det}).  

\section{GW Search Methods}
\label{sec:methods}
\subsection{C\lowercase{o}C\lowercase{o}A in brief}
\label{sec:cocoa}
\texttt{CoCoA} \citep{CoyneR+2016} is a GW search algorithm that adapts the cross-correlation technique originally developed for the detection of stochastic and continuous GW signals \citep{2008PhRvD..77h2001D,2022ApJ...941L..30A} to the search for GW transients of intermediate duration \citep{CoyneR+2016,2019PhRvD.100l4041S,2026PhRvD.113j3034K}, i.e., much shorter than continuous GWs (e.g.,\cite{Abbott+19_LongLivedBNS,ABbott+22_O3})
but longer than typical so-called GW burst searches (e.g., \cite{2005PhRvD..72d2002A,2025PhRvD.112j2005A,2024PhRvD.110d2007S}).

\texttt{CoCoA} searches are triggered GW searches aimed at discovering GW signals of intermediate duration from sources of known sky position occurring at a well-known trigger time (ideally with an uncertainty $\lesssim 1$\,s; \citep{2019PhRvD.100l4041S}). Searches also assume quasi-periodic GW signals  i.e., signals that can be considered
monochromatic over small time intervals, and whose time-frequency evolution can be modeled with sufficient physical accuracy over a time interval $T_{\rm coh}$, less than or equal to the total observation time $T_{\rm obs}$ over which the signal is expected to last. Since the signal is quasi-periodic, one can define a baseline $\Delta T_{\rm SFT} \le T_{\rm coh}$ such that, when performing a short Fourier Transform (SFT) of the signal within $T_I-\Delta T_{\rm SFT}$ to $T_I + \Delta T_{\rm SFT}$, all of the signal power at time $T_I$ is concentrated in a single SFT bin around a frequency $f(T_I)$, so the GW signal can be approximated as: 
\begin{eqnarray}
\nonumber h(t) \approx h_0(T_I){\cal A}_+F_+\cos(\phi(T_I)+2\pi f(T_I)(t-T_I))\\ + h_0(T_I){\cal A}_{\times}F_{\times}\sin(\phi(T_I)+2\pi f(T_I)(t-T_I)),~~~
\label{eq:amplitude-at-det}
\end{eqnarray}
where $F_+$ and $F_{\times}$ are the antenna factors that quantify the detector’s sensitivity to each GW polarization state, and where ${\mathcal A_+}$ and ${\mathcal A}_{\times}$ are amplitude factors dependent on the
physical system’s inclination angle $\iota$:
\begin{eqnarray}
{\mathcal A}_{+}=\frac{1+\cos^2{\iota}}{2}\\
{\mathcal A}_{\times}=\cos{\iota}.
\end{eqnarray}
Hereafter, we assume that $F_{+}$ and $F_{\times}$ are constant with time, given the duration of the signals we are interested in (see also \citep{CoyneR+2016}).

For each waveform, assuming a source located at a certain distance $d_L$, we can compute the root mean square (rms) amplitude at the detector as (see Equation 4.24 of \citep{CoyneR+2016}):
\begin{equation}
    h_{\rm rms} = \sqrt{\frac{\sum_{I} h_0^2(T_I)}{N_{\rm SFT}}},
    \label{eq:h_input}
\end{equation}
where $I$ is the SFT index; $N_{\rm SFT}=T_{\rm obs}/\Delta T_{\rm SFT}$; and $h_0(t)$ is proportional to the inverse of the distance to the source. 

With \texttt{CoCoA}, a cross-correlation statistic is calculated from SFTs of detector's data along the model time frequency tracks that form the search template bank. In the so-called stochastic limit of \texttt{CoCoA}, only SFT pairs from different detectors at the same time (after correcting for the time-of-flight in case of non co-located detectors) are correlated. This choice minimizes computational cost and maximizes robustness against signal uncertainties, at the expense of sensitivity. 
In the matched-filter limit of \texttt{CoCoA}, all possible SFT pairs (including self-pairs) are considered. This limit maximizes sensitivity at the expense of robustness and computing time. 
Finally, in the semi-coherent regime of \texttt{CoCoA}, a \texttt{CoCoA} matched-filter search is performed in each of  $N_{\rm coh} = T_{\rm obs} / T_{\rm coh}$  coherent segments, and the results for each coherent time segment are then combined incoherently. Hence, the \texttt{CoCoA} semi-coherent approach can be regarded as the incoherent sum of $N_{\rm coh}$ matched-filter searches carried out over $N_{\rm coh}$ time segments each of duration $T_{\rm coh}$. 

Hereafter, we focus on the \texttt{CoCoA} stochastic limit. In this limit, the detection cross-correlation statistic is a Gaussian variable with mean ($\mu_{\rho}$) and standard deviation ($\sigma_{\rho}$) defined as follows (see Equations 4.17 and 4.18 of \cite{CoyneR+2016}):
\begin{widetext}
\begin{equation}
     \mu_{\rho} = ({\mathcal A}_{+}^2 F^2_{+,H} + {\mathcal A}_{\times}^2 F^2_{\times,H})({\mathcal A}_{+}^2 F^2_{+,L} + {\mathcal A}_{\times}^2 F^2_{\times,L}) \frac{\Delta T^2_{SFT}}{2} \sum_{I}^{N_{SFT}} \frac{h_0^2(T_I)}{S_n^H [f_{k,I}] S_n^L [f_{k,I}] },
 \label{eq:mean_wideeq1}
\end{equation}
\end{widetext}
\begin{widetext}
\begin{equation}
     \sigma_{\rho}^2 = (\mathcal{A}_{+}^2 F^2_{+,H} + \mathcal{A}_{\times}^2 F^2_{\times,H})(\mathcal{A}_{+}^2 F^2_{+,L} + \mathcal{A}_{\times}^2 F^2_{\times,L}) \frac{\Delta T^2_{SFT}}{2} \sum_{I}^{N_{SFT}} \frac{1}{S_n^H [f_{k,I}] S_n^L [f_{k,I}]},
 \label{eq:rho_wideeq2}
\end{equation}
\end{widetext}
where $S_n$ is the single-sided Power Spectral Density (PSD)
of the noise; the $H$ and $L$ subscripts denote quantities calculated for two different detectors, i.e., in this example, LIGO Hanford and LIGO Livingston, respectively. Assuming white Gaussian noise (so that $S_n$ is independent of frequency), the detection condition can be expressed as $h_{\rm rms} (1/d) \gtrsim  h_{\rm stoch}$ with (see Equation 4.23 in \citep{CoyneR+2016}):
\begin{equation}
    h_{\rm stoch}=\frac{\sqrt{2}\mathcal{S}^{1/2} \Delta T_{SFT}^{-1/2}N_{SFT}^{-1/4}{(S_n^HS_n^L)}^{1/4}}{[(\mathcal{A}^2_+ F^2_{+,H} + \mathcal{A}^2_\times F^2_{\times,H})(\mathcal{A}^2_+ F^2_{+,L} + \mathcal{A}^2_\times F^2_{\times,L})]^{1/4}},
    \label{eq:h_min_stoch}
\end{equation}
where ${\mathcal S}$ is defined as:
\begin{equation}
    {\mathcal S}={\rm erfc}^{-1}({2\alpha})-{\rm erfc}^{-1}({2\gamma}),
\end{equation}
where $\alpha$ is the False Alarm Probability (FAP), $1-\gamma$ is the False Dismissal Probability (FDP), ${\rm erfc}(x)$ is the complementary error function of $x$, and ${\rm erfc}^{-1}(x)$ is its inverse. The complementary error function is related to the error function by the relation ${\rm erfc}(x) = 1-{\rm erf}(x)$, where the ${\rm erf}(x)$ is the probability that a random variable that is normally distributed with mean zero and standard deviation $1/\sqrt{2}$ falls in the range $[-x, x]$. Hence, the complementary error function represents the area under the two tails of a zero-mean
Gaussian probability density function with standard deviation $1/\sqrt{2}$. 

\subsection{EM- \lowercase{vs} GW-triggered search strategies}
\label{sec:strategies}
Based on the tentative association between S250818k and SN\,2025ulz, we envisage two types of \texttt{CoCoA} GW search strategy for sub-solar NSs in the context of the superkilonova scenario, described in what follows.

The first is inspired directly by the S250818k/SN\,2025ulz potential association: a low-significance matched-filter GW trigger provides an initial sky localization area, within which an EM campaign searches for counterparts, ideally yielding a much more accurate sky position through the identification of a SN or other EM counterpart \citep{KasliwalM+25_S250818k,AckleyK+2026}. A more robust \texttt{CoCoA} search is then carried out using the GW-provided merger time and the EM-provided sky position. We stress, however, that the GW sky localization is likely to be large when derived from a low-threshold matched-filter trigger, so our ability to conduct a \texttt{CoCoA} search would rely on identifying an EM counterpart within that area.

The second type of search is independent of the GW localization capabilities and is usually referred to as an EM-triggered search (e.g., \citep{2022ApJ...928..186A,PhysRevD.101.084002, 2024PhRvD.110d2007S,2025ApJ...985..183A}). Here one would use \texttt{CoCoA} to search for GWs within an on-source time window and sky localization area both provided by independent EM observations (see e.g., \citep{2025A&A...701A.128A,2025A&A...701A.128A,ODwyerT+26,2016ApJ...830...42C}). Indeed superkilonovae, if they exist, are potentially identifiable through spectroscopic follow-up of long GRBs or engine-driven stripped-envelope SNe (such as Type IIb or broad-lined Type Ic; \citealp{Barnes&Metzger22,Siegel2021}, and Section \ref{sec:rates}). They are generally expected to be even redder than kilonovae owing to line blanketing by the $r$-process elements produced. The near- and mid-infrared are of particular interest, as this is the regime where neutron-capture ($r$-process) lines can be identified. Alternatively, superkilonovae could be identified using the wide-field capabilities of the upcoming Roman \citep{Springel2015} and/or Rubin LSST \citep{2019ApJ...873..111I,Tyson2002} observatories. Their light curves and spectra are distinguishable from those of ordinary Type I and II SNe, and theory predicts that they should be detectable out to $z\approx 1$ with Roman NIR observations reaching a depth of 26-27\,mag in F158, F184, and F62 filters \citep{RoseB+21,Siegel2021}. An EM-triggered GW search for sub-solar NS-BH mergers with \texttt{CoCoA} would benefit from a very small sky localization error, but its sensitivity is likely to be limited by how well the merger time of the sub-solar NS-BH system can be reconstructed from the EM observations alone, since the larger the timing uncertainty, the less sensitive GW searches become (e.g., \citep{2019PhRvD.100l4041S,2024PhRvD.110d2007S}).

\begin{figure}
    \centering
    \includegraphics[width=\linewidth]{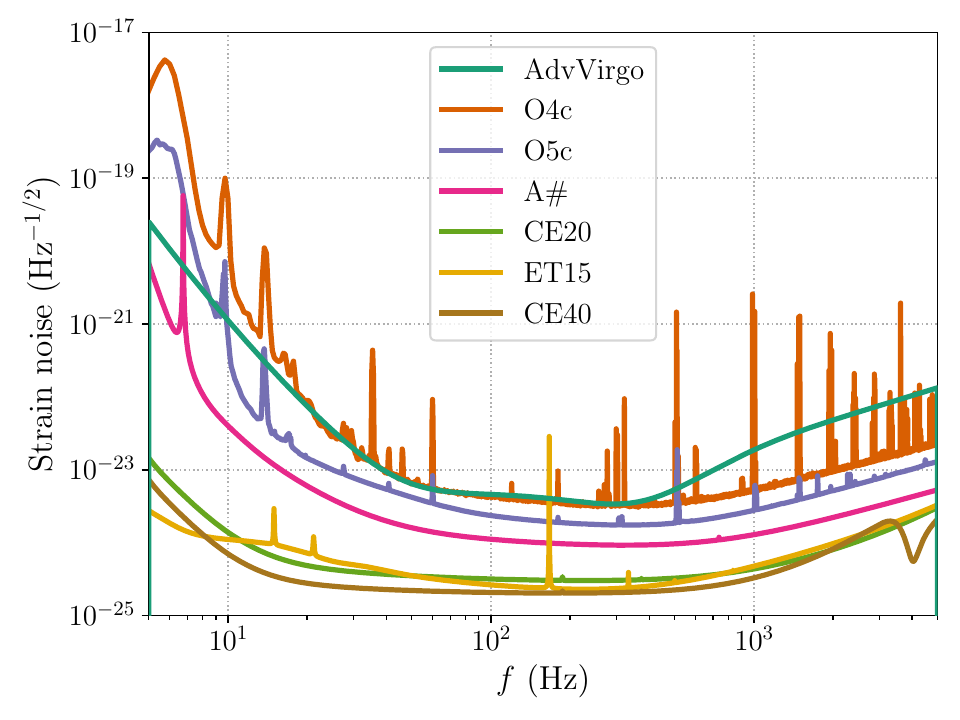}
    \caption{Amplitude spectral density (ASD) curves for all the GW detectors used in this study (Section~\ref{sec:det}). While for our localization results shown in Figure \ref{fig:loc} we use the respective L1 and H1 ASDs for O4c as given in Ref.~\cite{O4c_sensitivity}, since they are very similar, we only show O4c for LIGO-H here for clarity.}
    \label{fig:psds}
\end{figure}

\section{Detector sensitivity and networks}
\label{sec:det}
We assess the detection and localization capabilities of current- and next-generation (XG) GW detectors in different network configurations. The noise curves (noise amplitude spectral density; ASD) used for the distance range estimates described in Section \ref{sec:results} are shown in Figure~\ref{fig:psds}. Specifically, to provide the range of distance horizons for a single-trial one- or two-detector searches (Section \ref{sec:Gwhorizons}), we consider the LIGO Hanford detector (H) at O4c sensitivity~\citep{O4c_sensitivity, PhysRevD.111.062002} and a 40 km Cosmic Explorer (CE40) at default sensitivity \cite{CE_sensitivity, 2021arXiv210909882E} plus a 15\,km Einstein Telescope in L--configuration \citep[ET-L1; ][]{ET_sensitivity_CoBA,ET:2025xjr, 2010CQGra..27h4007P}. We stress that all the sensitivity curves used in this study are taken from published and publicly available ASDs. In addition, we consider GW localization capabilities for the following detector networks:
\begin{enumerate}
    \item O4 network: LIGO Livingston (L), H, and Virgo (V) detectors with LIGO detectors taken at O4c sensitivity~\citep{O4c_sensitivity, PhysRevD.111.062002, 2018LRR....21....3A} and Virgo at Advanced Virgo design sensitivity~\citep{2015CQGra..32b4001A}.
    \item O5 network: L, H, and V detectors with LIGO detectors taken at O5c/A+ sensitivity~\citep{O5c_sensitivity, 2018LRR....21....3A} and Virgo at Advanced Virgo design sensitivity~\citep{2015CQGra..32b4001A}.
    \item A\# network: L, H, and LIGO-India \citep{Iyer2011_LIGOIndia} (I) detectors, all at A\# sensitivity~\citep{ASharp_sensitivity, 2018LRR....21....3A}.
    \item XG network: CE40 at default sensitivity in \cite{CE_sensitivity}, a 20 km Cosmic Explorer (CE20) at default sensitivity for CE20 in \cite{CE_sensitivity}, and two 15\,km Einstein Telescopes (ET-L1, ET-L2) in L--configuration at ET15 sensitivity~\citep{ET_sensitivity_CoBA} used in the CoBA study~\citep{ET:2025xjr}.
\end{enumerate}

\begin{table}
\centering
\def\arraystretch{1.3}
\setlength{\tabcolsep}{4pt}
\caption{Detector locations and networks included in this study. \label{tab:networks}}
\begin{tabular}{ccccc}
\hline
\hline
Label & Latitude      & Longitude       & Azimuth        & Network \\ \hline
L     & $30.56^\circ$ & $-90.77^\circ$  & $197.7^\circ$  & O4, O5, A\#        \\ \hline
H     & $46.46^\circ$ & $-119.4^\circ$  & $126^\circ$    & O4, O5, A\#        \\ \hline
V    & $43.63^\circ$ & $10.50^\circ$   & $19.43^\circ$  & O4/O5              \\ \hline
I     & $19.61^\circ$ & $77.03^\circ$   & $117.62^\circ$ & A\#        \\ \hline
CE40  & $34.17^\circ$ & $-118.13^\circ$ & $90^\circ$     & XG              \\ \hline
CE20  & $29^\circ$    & $-94^\circ$     & $-200^\circ$   & XG              \\ \hline
ET-L1    & $40.52^\circ$ & $9.42^\circ$   & $0^\circ$  & XG              \\ \hline
ET-L2    & $50.72^\circ$ & $5.92^\circ$   & $45^\circ$  & XG              \\ \hline
\end{tabular}
\label{tab:det-loc}
\end{table}

\begin{table}
\def\arraystretch{1.2}
\setlength\tabcolsep{20pt}
\centering
\caption{Antenna pattern values used for the GW sensitivity (horizon distance) estimates. We assume a face-on optimally-oriented system for the LIGO Hanford (LIGO-H) detector with (R.A., Dec., $\psi$) = (-0.2572, 0.8108, 0)\,rad, $\iota=0$~rad, and $t_\text{GPS}=0$~s. Columns from left to right are: Detector symbols, $F_+$, and $F_\times$. }
\begin{tabular}{ccc}
\hline
\hline
Detector & $F_+$ & $F_{\times}$ \\ 
\hline
L & 0.2744 & -0.8476 \\ 
H & -0.3090 & 0.9510 \\ 
V & 0.2079 & 0.0517 \\ 
I & 0.5641 & -0.1570 \\ 
CE40 & -0.9768 & 0.0284 \\ 
CE20 & -0.8877 & -0.1308 \\ 
ET-L1 & 0.0649 & -0.2384 \\ 
ET-L2 & 0.2992 & 0.4381 \\ 
\hline
\end{tabular}
\label{tab:fp-fc}
\end{table}

The locations of all the detectors considered in these networks are given in Table~\ref{tab:det-loc}.

\section{Results}
\label{sec:results}

\subsection{GW distance horizons}
\label{sec:Gwhorizons}
Here, we estimate the distance horizons for sub-solar NS-BH single-trial one-detector matched-filter searches and single-trial two-detector \texttt{CoCoA} stochastic searches. In each case we assume the minimum number of detectors required to implement the search. Real searches typically rely on networks of two or more detectors,  improving sensitivity and potentially providing GW localization, and use large template banks, resulting in a correspondingly large number of trials (e.g., ${\cal O}(10^6)$ trials were estimated for a two-parameter-waveform \texttt{CoCoA} EM-triggered search by \citet{2019PhRvD.100l4041S}). Our aim here, however, is to characterize the distance reach of the most sensitive technique (matched filtering) and of a more robust one (the \texttt{CoCoA} stochastic search) under these minimal assumptions. The resulting distances provide a useful reference, as they can be rescaled to obtain order-of-magnitude sensitivity estimates for more complex and realistic searches.
For all distance estimates, we consider an optimally-oriented source with respect to LIGO Hanford (see Table\,\ref{tab:fp-fc}). 

For the 15 LAL waveforms described in Section\,\ref{sec:lalwave}, we report in Table \ref{tab:lal_sensitivity} the optimal SNR ($\rho^{\rm opt}$) at 112\,Mpc, and the optimal horizon distance estimated as the distance at which $\rho^{\rm opt}\approx 8$ (i.e., single-trial False-Alarm-Probability of FAP$\approx 1.3\times10^{-14}$) in a single-detector matched-filter search with a LIGO H-like detector or a 40-km CE detector with ideal Gaussian noise. The optimal horizons should be seen as the most optimistic estimates of the reach of single-detector searches for sub-solar NS-BH systems. We have verified that these horizon distance estimates are fully consistent with what one would derive using the analytical waveforms described in Section\,\ref{sec:analyticalwave}. 

\begin{figure*}
    \centering
    \includegraphics[width=\linewidth]{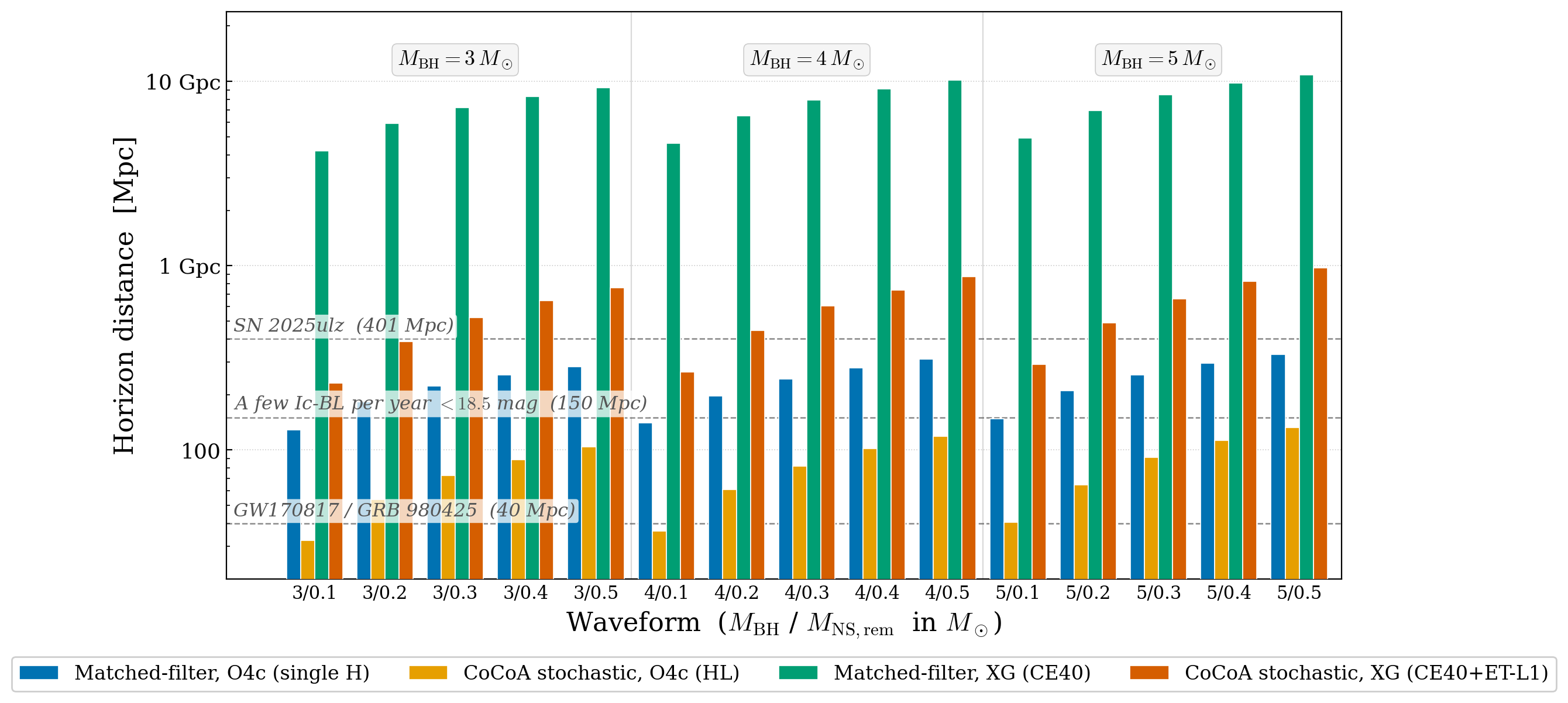}
    \caption{Subsolar waveform GW horizons across the search strategies analyzed in this work. See Section \ref{sec:Gwhorizons} for discussion.}
    \label{fig:histo}
\end{figure*}

Next, using the formulation described in Section \ref{sec:cocoa}, we estimate the \texttt{CoCoA} stochastic horizon distance for a  single-trial FAP$\approx 1.3\times10^{-14}$ and FDP\,$\approx 50\%$, assuming a two-detector LIGO-H plus LIGO-L search with O4c-like ASD and Gaussian noise. For the \texttt{CoCoA} search we set $\Delta T_{\rm SFT} = 9$\,ms to satisfy the quarter-cycle approximation. Finally, we repeat the \texttt{CoCoA} estimates for a two-detector stochastic search with CE40 and one ET 15\,km (ET-L1). We note that the \texttt{CoCoA} horizon distances of an actual search may be degraded to up to a factor of $\approx \sqrt{2}$ in distance reach compared to what one would estimate purely analytically (Equation \ref{eq:h_min_stoch}), due to spectral leakage effects \citep{CoyneR+2016}. Because spectral leakage is equivalent to a systematic error, it will manifest itself more prominently in XG detectors. For this reason, the estimates reported in Table \ref{tab:lal_sensitivity} are obtained applying a Hanning window to the signal (in a zero-noise approximation) and then performing actual SFTs on the signal to calculate the expectation value (see the numerator in the sum of Equation \ref{eq:mean_wideeq1}). The detector ASD is then used to weigh the signal power in each SFT and frequency bin, as in Equation \ref{eq:mean_wideeq1}, and to calculate the standard deviation of the detection statistic as in Equation \ref{eq:rho_wideeq2}.

As evident from Table \ref{tab:lal_sensitivity} and Figure\,\ref{fig:histo}, \texttt{CoCoA} stochastic searches in current-generation GW detectors data may start setting interesting constraints on sub-solar NS mergers in the superkilonova scenario via EM-triggered strategies. Indeed, the distance horizons in Table \ref{tab:lal_sensitivity} are at least comparable to the distances of the closest short (GRB170817/GW170817) and long (SN\,1998bw/GRB980425) GRBs we know of, both located at $d_L\approx 40$\,Mpc \citep{1998Natur.395..670G,2017ApJ...848L..31H}. While these events are very rare (order of one per decade or less), they have occurred. Further sensitivity improvements in current-generation GW detectors of a factor of $\approx 1.5$ compared to the O4 sensitivity, such as the planned A\#, would be very helpful especially for ensuring that the horizon distances of stochastic searches with \texttt{CoCoA} get closer to $\approx 150$\,Mpc: this is the distance at which optical surveys of the transient sky are currently detecting and spectroscopically classifying a few Ic-BL SNe per year (e.g., \cite{ODwyerT+26,2024ApJ...976...71S}). In order for \texttt{CoCoA} searches to reach distances comparable to that of SN\,2025ulz, which was located at $\approx 401$\,Mpc (e.g., \citep{KasliwalM+25_S250818k,GillandersH+25_PanStarrs}), next-generation GW detectors such as CE and ET are going to be critical. Using the superkilonova rate upper-limit discussed in Section \ref{sec:rates} (see Eq. \ref{eq:SKNrate}), we estimate $\lesssim 85\%$ probability of detecting at least one superkilonova at $\lesssim 400$\,Mpc over a 1-yr period. 
As we show in the next Section, these detectors are also expected to provide  much improved localizations for GW-triggered searches with \texttt{CoCoA}.

\begin{table*}
\centering
\caption{Sub-solar waveform detectability estimates (see Section \ref{sec:Gwhorizons} for details) for the 15 LAL waveforms described in Section \ref{sec:lalwave}. The first six columns, from left to right are: central BH mass; NS merger-remnant mass;  in-band duration (time to span 20--1024\,Hz); optimal SNR at 112\,Mpc; optimal horizon distance (SNR$\approx 8$, FAP$\approx 1.3\times10^{-14}$) for a single-detector matched-filter search with a LIGO H-like detector, Gaussian noise, and an O4c-like ASD; and \texttt{CoCoA} stochastic horizon distance (single-trial FAP$\approx 1.3\times10^{-14}$, FDP$\approx 50\%$) for a 2-detector LIGO HL search with O4c-like ASD and Gaussian noise. The last three  columns are similar estimates but for CE40\,km and ET-L1. For the \texttt{CoCoA} horizons we set $\Delta T_{\rm SFT} = 9$\,ms so as to satisfy the quarter-cycle approximation. }
\label{tab:lal_sensitivity}
\begin{tabular}{ccccccccc}
\hline
$M_{\rm BH}$ & $M_{\rm NSrem}$ & T$_{\rm 20-1024\,Hz }$  & $\rho^{\rm opt,O4c}_{\rm H,112\,Mpc}$ &  $d^{\rm opt,O4c}_{\rm H,1-trial}$& $d^{\rm CoCoA-S,O4c}_{\rm HL,1-trial}$ & $\rho^{\rm opt,XG}_{\rm CE40,112\,Mpc}$ &  $d^{\rm opt,XG}_{\rm CE40,1-trial}$& $d^{\rm CoCoA-S,XG}_{\rm CE40-ETL1,1-trial}$\\
$(M_\odot)$ & $ (M_\odot)$ & (s)   &  & (Mpc) & (Mpc) & & (Gpc) & (Mpc) \\
\hline
3.0 & 0.1 &   1078 &  9.23   & $129$ & 32.2 & 301 & 4.21& 231\\
3.0 & 0.2 &    545&  13.0   & $182$ & 53.7 & 423 & 5.92& 387\\
3.0 & 0.3 &   368 &   15.9   & 223& 72.6 & 516 & 7.22& 522 \\
3.0 & 0.4 &   279 &   18.2    & 255& 88.5& 593 & 8.30& 648\\
3.0 & 0.5 &   225  &  20.3  & 284 & 104& 661 & 9.25& 759\\
\hline
4.0 & 0.1 &   887 &   10.0 & 140& 36.4& 329 & 4.61& 266\\
4.0 & 0.2 &   447 &  14.1  & 197& 61.2&  464 & 6.50& 445\\
4.0 & 0.3 &   301 &  17.3 & 242& 82.0& 566 & 7.92& 604\\
4.0 & 0.4 &   227 & 19.9 & 279& 102& 651 & 9.11& 740\\
4.0 & 0.5 &   183 &  22.2& 311& 119& 726 & 10.2& 872\\
\hline
5.0 & 0.1 &    762&   10.6 & 148   & 40.5& 352 & 4.93& 292\\
5.0 & 0.2 &   384 &  15.0 & 210  & 64.8& 496 &6.94& 489\\
5.0 & 0.3 &   258&   18.3 & 256   & 91.4& 607 & 8.50&  659\\
5.0 & 0.4 &   195 &   21.1 & 295  &  113& 699 & 9.79& 823\\
5.0 & 0.5 &  157 &   23.6 & 330 &132 & 779 & 10.9 & 973\\
\hline
\end{tabular}
\end{table*}

\subsection{GW localization for GW-triggered searches}
To evaluate the feasibility of GW-triggered searches for sub-solar NS-BH systems (see Section \ref{sec:strategies}), we generate a population of non-spinning sub-solar NS-BH systems similar to that described in Section \ref{sec:lalwave} and estimate the localization capabilities of various GW detector networks. Specifically, we sample 500 points uniformly in chirp mass such that the primary mass lies between $m_1 \in [3, 5]~M_\odot$ and the secondary mass lies between $m_2 \in [0.2, 1]~M_\odot$. We then distribute the simulated systems in the sky, adopting a uniform R.A. distribution in the $[0, 2\pi]$ range, and a uniform-Cosine distribution for the Dec in the range $[-1, 1]$. We also adopt a uniform distribution in polarization angle ($\psi$) in the range $[0, \pi]$. We place all the binaries at a distance of $d_L=112$~Mpc in face-on configuration ($\iota=0$). We have chosen a GW phase equal to zero at $t_\text{gps}=0$.

\begin{figure}
    \centering
    \includegraphics[width=\linewidth]{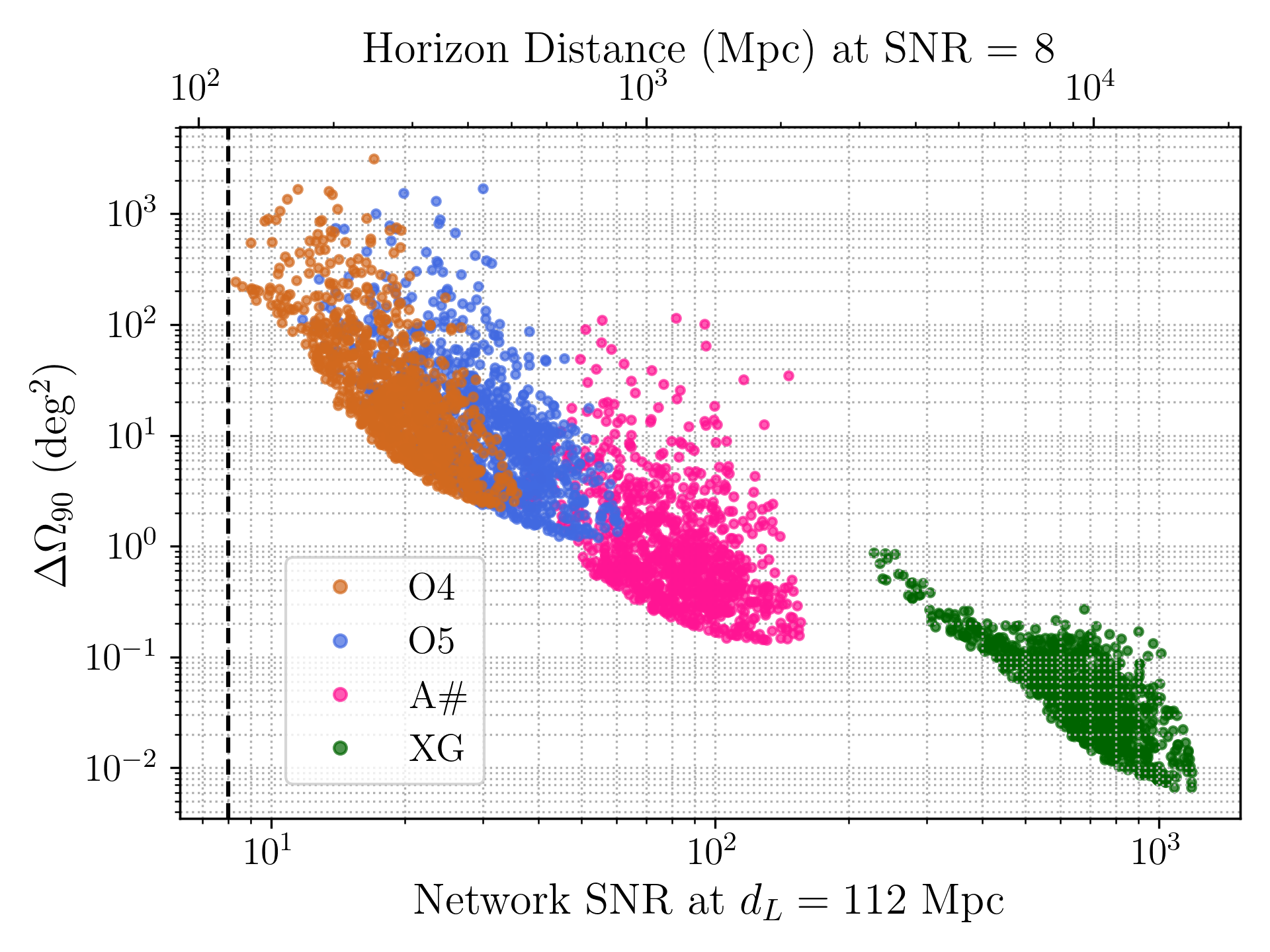}
    \caption{90\% Localization area (deg$^2$) as a function of network SNR for various detector networks (see Section \ref{sec:det} for details). Networks of three current generation detectors at their O4 (orange), O5/A+ (blue), and A\# (magenta) sensitivity are compared with a network of XG detectors (green). All binaries are placed at $d_L=112$\,Mpc. The upper axis shows, for each network, the distance that would yield a network SNR of 8. For reference, the 90\% GW sky localization area of S250818k was $\approx 949$\,deg$^2$ \citep{GCN41440}.}
    \label{fig:loc}
\end{figure}

For these sources, we generate signals using the waveform model \textsc{IMRPhenomXPHM}~\citep{Pratten:2020ceb, lalsuite}\footnote{Here, we use a more advanced waveform model (\textsc{IMRPhenomXPHM}), which includes higher-modes, compared to \textsc{IMRPhenomD} (used in previous sections) because higher-mode waveforms are better suited for parameter estimation and localization calculations since they help in breaking degeneracies between different parameters such as luminosity distance and inclination angle.} and calculate the optimal SNR and sky localization area using the timing-based triangulation method \citep{Fairhurst:2009tc, Fairhurst:2017mvj, Fairhurst:2023idl} (see Appendix B.2. of Fairhurst et al.~\cite{Fairhurst:2017mvj} for details) for O4, O5, A\#, and XG networks listed in Table\,\ref{tab:networks}. In Figure~\ref{fig:loc} we show the 90\% GW sky localization area for binaries with SNR $> 5$ in each detector and a network SNR $> 8$ when placed at $d_L=112$\,Mpc. While the estimates performed here should be taken with caution as they represent  the best possible localizations achievable by a given detector network as they assume perfect waveform match, it can be seen that the XG network performs better by at least two orders of magnitude compared to the current-generation detector networks at O4 sensitivity. Hence, with next-generation detectors, the search for EM counterparts to sub-solar NS-BH systems, and the corresponding GW-triggered searches for long-duration GWs, will be greatly facilitated.

\section{Summary and Conclusion}
\label{sec:conclusion}
Prompted by the tentative association between the sub-threshold candidate S250818k and the Type~IIb supernova SN~2025ulz, in this work we have addressed the question of whether it is possible to detect long-lived, non-standard GW chirps potentially associated with them using existing and planned GW instruments. Because the astrophysics of sub-solar NS formation and the subsequent merger with lower-mass-gap BHs can push the waveforms away from reliable templates, we considered detection techniques beyond matched filtering, focusing on the more robust but less sensitive cross-correlation algorithm \texttt{CoCoA}, applied either as follow-up of candidate chirps or as EM-triggered searches of stripped-envelope CCSNe. This is the first analysis where \texttt{CoCoA} has been applied to a class of waveforms substantially different from the bar-mode instability waveforms for which \texttt{CoCoA} was originally proposed \citep{CoyneR+2016,2019PhRvD.100l4041S}.

The central conclusion of our work is that EM-triggered stochastic searches for sub-solar chirps are worthwhile in current-generation GW detector data for nearby events: the expected reach covers the distances at which the nearest GW/GRB-associated explosions have historically occurred, though the expected rate of such events is low. The value of an A\# upgrade on current-generation detectors is clear, as it moves the accessible volume up to where transient surveys routinely classify the relevant SN population, turning opportunistic follow-up into more of a systematic program. Reaching the population as a whole, however, is fundamentally in the realm of next-generation GW detectors: only CE and ET extend the horizon to the distances of events like SN\,2025ulz, and could deliver the localizations that make GW-triggered searches practical alongside EM-triggered ones. Our rate-based estimate reinforces this conclusion, indicating that a confident yearly detection is plausible only once sensitivities approach several hundred Mpc. This conclusion is affected by a big uncertainty, namely, the current lack of a robust rate measurement for superkilonovae. The last will require a large sample of confident subsolar-mass GW detections with well-characterized GRB or SN counterparts, together with global simulations of the full disk evolution.

We stress that the superkilonova scenario itself remains unconfirmed; S250818k is sub-threshold, and the link to SN\,2025ulz is suggestive rather than established. Our results should thus be read as a detectability roadmap for a hypothesis worth testing, not as evidence for it. Moreover, \texttt{CoCoA}'s robustness comes at a sensitivity cost, so the horizons quoted here are conservative compared to a matched-filter search (Table \ref{tab:lal_sensitivity}) and driven by the astrophysical uncertainties. Any future narrowing of the theoretical scenarios invoking superkilonovae, and of the related GW signal models, would therefore help extend GW searches to larger distances. In fact, in both the matched filter and \texttt{CoCoA} cases we have considered only the single-trial horizons. These are useful for projecting sensitivities in a variety of scenarios that could include different search strategies (Section \ref{sec:strategies}) and/or GW networks (Section \ref{sec:det}). However, a real search would need to account for the number of trials namely, the size of the template bank, that is likely to be dominated by the uncertainties on the expected GW signals (e.g., on the component masses) and, in an EM-triggered framework, by the uncertainties on the merger time \cite{2019PhRvD.100l4041S}. The number of trials also impacts the computing cost, which would need to be carefully evaluated in a search where computing resources may be limited \citep{2026PhRvD.113j3034K} and/or computing speed is of essence (e.g., low-latency searches).

\begin{acknowledgements}
\small
A.C. acknowledges support from the National Science Foundation (NSF) via the grant \# AST-2431072. DJ acknowledges the Science and Technology Facilities Council (STFC) for support through grants ST/V005618/1 and ST/Y004272/1. YZ acknowledges support from MAOF grant 12641898 and visitor support from the Observatories of the Carnegie Institution for Science, Pasadena CA, where much of this work was completed. RC acknowledges support from the NSF via grant \# PHY-2512922.

\end{acknowledgements}

\bibliographystyle{aasjournal}
\bibliography{main}

@ARTICLE{GCN41440,
       author = {{Ligo Scientific Collaboration} and {VIRGO Collaboration} and {Kagra Collaboration}},
        title = "{LIGO/Virgo/KAGRA S250818k: Updated Sky localization and EM Bright Classification}",
      journal = {GRB Coordinates Network},
         year = 2025,
        month = aug,
       volume = {41440},
        pages = {1},
       adsurl = {https://ui.adsabs.harvard.edu/abs/2025GCN.41440....1L}
}

@ARTICLE{2026arXiv260505444T,
       author = {{Abac}, A.~G. and {Abouelfettouh}, I. and {Acernese}, F. and {Ackley}, K. and {Adam}, A. and {Adamcewicz}, C. and {Adhicary}, S.},
        title = "{Searches for Binary Mergers with Sub-solar Mass Components in Data from the First Part of LIGO--Virgo--KAGRA's Fourth Observing Run}",
      journal = {arXiv e-prints},
         year = 2026,
        month = may,
          eid = {arXiv:2605.05444},
        pages = {arXiv:2605.05444},
          doi = {10.48550/arXiv.2605.05444},
archivePrefix = {arXiv},
       eprint = {2605.05444},
 primaryClass = {astro-ph.HE},
       adsurl = {https://ui.adsabs.harvard.edu/abs/2026arXiv260505444T}
}

@article{PhysRevLett.123.161102,
  title = {Search for Subsolar Mass Ultracompact Binaries in Advanced LIGO's Second Observing Run},
  author = {Abbott, B. P. and Abbott, R. and Abbott, T. D. and Abraham, S. and Acernese, F. and Ackley, K. and Adams, C. and Adhikari, R. X.},
  collaboration = {LIGO Scientific Collaboration and the Virgo Collaboration},
  journal = {Phys. Rev. Lett.},
  volume = {123},
  issue = {16},
  pages = {161102},
  numpages = {13},
  year = {2019},
  month = {Oct},
  publisher = {American Physical Society},
  doi = {10.1103/PhysRevLett.123.161102},
  url = {https://link.aps.org/doi/10.1103/PhysRevLett.123.161102}
}

@article{PhysRevLett.129.061104,
  title = {Search for Subsolar-Mass Binaries in the First Half of Advanced LIGO's and Advanced Virgo's Third Observing Run},
  author = {Abbott, R. and Abbott, T. D. and Acernese, F. and Ackley, K. and Adams, C. and Adhikari, N. and Adhikari, R. X.},
  collaboration = {LIGO Scientific Collaboration and Virgo Collaboration},
  journal = {Phys. Rev. Lett.},
  volume = {129},
  issue = {6},
  pages = {061104},
  numpages = {16},
  year = {2022},
  month = {Aug},
  publisher = {American Physical Society},
  doi = {10.1103/PhysRevLett.129.061104},
  url = {https://link.aps.org/doi/10.1103/PhysRevLett.129.061104}
}

@article{LVK_subsolar_O3b,
    author = {{The LIGO Scientific Collaboration} and {the Virgo Collaboration} and {the KAGRA Collaboration}},
    title = {Search for subsolar-mass black hole binaries in the second part of Advanced LIGO's and Advanced Virgo's third observing run},
    journal = {Monthly Notices of the Royal Astronomical Society},
    volume = {524},
    number = {4},
    pages = {5984-5992},
    year = {2023},
    month = {10},
    issn = {0035-8711},
    doi = {10.1093/mnras/stad588},
    url = {https://doi.org/10.1093/mnras/stad588},
}

@article{83j3-pgk1,
  title = {All-sky search for long-duration gravitational-wave transients in the first part of the fourth LIGO-Virgo-KAGRA observing run},
  author = {Abac, A. G. and Abouelfettouh, I. and Acernese, F. and Ackley, K. and Adamcewicz, C. and Adhicary, S. and Adhikari, D.},
  collaboration = {LIGO Scientific Collaboration, Virgo Collaboration, and KAGRA Collaboration},
  journal = {Phys. Rev. D},
  volume = {113},
  issue = {8},
  pages = {082004},
  numpages = {23},
  year = {2026},
  month = {Apr},
  publisher = {American Physical Society},
  doi = {10.1103/83j3-pgk1},
  url = {https://link.aps.org/doi/10.1103/83j3-pgk1}
}

@article{2022ApJ...941L..30A,
    author = {{Abbott}, R. and {Abe}, H. and {Acernese}, F. and et al. and
              {The LIGO Scientific Collaboration} and {the Virgo Collaboration}
              and {the KAGRA Collaboration}},
    title = {Model-based Cross-correlation Search for Gravitational Waves from
             the Low-mass X-Ray Binary Scorpius X-1 in LIGO O3 Data},
    journal = {The Astrophysical Journal Letters},
    doi = {10.3847/2041-8213/aca1b0},
    url = {https://doi.org/10.3847/2041-8213/aca1b0},
    year = {2022},
    month = {dec},
    volume = {941},
    number = {2},
    pages = {L30},
}

@article{PhysRevD.101.084002,
  title = {Optically targeted search for gravitational waves emitted by core-collapse supernovae during the first and second observing runs of advanced LIGO and advanced Virgo},
  author = {Abbott, B. P. and Abbott, R. and Abbott, T. D. and Abraham, S. and Acernese, F. and Ackley, K. and Adams},
  collaboration = {LIGO Scientific Collaboration and Virgo Collaboration and ASAS-SN Collaboration and DLT40 Collaboration},
  journal = {Phys. Rev. D},
  volume = {101},
  issue = {8},
  pages = {084002},
  numpages = {24},
  year = {2020},
  month = {Apr},
  publisher = {American Physical Society},
  doi = {10.1103/PhysRevD.101.084002},
  url = {https://link.aps.org/doi/10.1103/PhysRevD.101.084002}
}

@ARTICLE{GWTC5_LIGO2026,
       author = {{Abac}, A.~G. and {Abe}, A. and {Abouelfettouh}, I. and {Acernese}, F. and {Ackley}, K. and {Adam}, A. and {Adhicary}, S.},
        title = "{GWTC-5.0: Population Properties of Merging Compact Binaries}",
      journal = {arXiv e-prints},
         year = 2026,
        month = may,
          eid = {arXiv:2605.27226},
        pages = {arXiv:2605.27226},
          doi = {10.48550/arXiv.2605.27226},
archivePrefix = {arXiv},
       eprint = {2605.27226},
 primaryClass = {astro-ph.HE},
       adsurl = {https://ui.adsabs.harvard.edu/abs/2026arXiv260527226T}
}

@ARTICLE{Sub_BNSLVK26,
       author = {{Abac}, A.~G. and {Abouelfettouh}, I. and {Acernese}, F. and {Ackley}, K. and {Adam}, A. and {Adamcewicz}, C. and {Adhicary}, S.},
        title = "{Searches for Binary Mergers with Sub-solar Mass Components in Data from the First Part of LIGO--Virgo--KAGRA's Fourth Observing Run}",
      journal = {arXiv e-prints},
         year = 2026,
        month = may,
          eid = {arXiv:2605.05444},
        pages = {arXiv:2605.05444},
          doi = {10.48550/arXiv.2605.05444},
archivePrefix = {arXiv},
       eprint = {2605.05444},
 primaryClass = {astro-ph.HE},
       adsurl = {https://ui.adsabs.harvard.edu/abs/2026arXiv260505444T}
}

@article{Cornish:2026ltu,
    author = {Cornish, Nelson Christensen. Neil and K{\"u}hnel, Florian and Sakellariadou, Mairi and Guanga, Andres Santiago Villares},
    title = "{Parameter Estimation on LIGO-Virgo-KAGRA O4a Binary Merger Triggers with Sub-solar Mass Components}",
    eprint = "2607.23119",
    archivePrefix = "arXiv",
    primaryClass = "gr-qc",
    month = "7",
    year = "2026"
}

@ARTICLE{2017ApJ...848L..31H,
       author = {{Hjorth}, Jens and {Levan}, Andrew J. and {Tanvir}, Nial R. and {Lyman}, Joe D. and {Wojtak}, Rados{\l}aw and {Schr{\o}der}, Sophie L. and {Mandel}, Ilya and {Gall}, Christa and {Bruun}, Sofie H.},
        title = "{The Distance to NGC 4993: The Host Galaxy of the Gravitational-wave Event GW170817}",
      journal = {\apjl},
         year = 2017,
        month = oct,
       volume = {848},
       number = {2},
          eid = {L31},
        pages = {L31},
          doi = {10.3847/2041-8213/aa9110},
archivePrefix = {arXiv},
       eprint = {1710.05856},
 primaryClass = {astro-ph.GA},
       adsurl = {https://ui.adsabs.harvard.edu/abs/2017ApJ...848L..31H}
}

@ARTICLE{1998Natur.395..670G,
       author = {{Galama}, T.~J. and {Vreeswijk}, P.~M. and {van Paradijs}, J. and {Kouveliotou}, C. and {Augusteijn}, T. and {B{\"o}hnhardt}, H. and {Brewer}, J.~P. and {Doublier}, V. and {Gonzalez}, J.-F. and {Leibundgut}, B. and {Lidman}, C. and {Hainaut}, O.~R. and {Patat}, F. and {Heise}, J. and {in't Zand}, J. and {Hurley}, K. and {Groot}, P.~J. and {Strom}, R.~G. and {Mazzali}, P.~A. and {Iwamoto}, K. and {Nomoto}, K. and {Umeda}, H. and {Nakamura}, T. and {Young}, T.~R. and {Suzuki}, T. and {Shigeyama}, T. and {Koshut}, T. and {Kippen}, M. and {Robinson}, C. and {de Wildt}, P. and {Wijers}, R.~A.~M.~J. and {Tanvir}, N. and {Greiner}, J. and {Pian}, E. and {Palazzi}, E. and {Frontera}, F. and {Masetti}, N. and {Nicastro}, L. and {Feroci}, M. and {Costa}, E. and {Piro}, L. and {Peterson}, B.~A. and {Tinney}, C. and {Boyle}, B. and {Cannon}, R. and {Stathakis}, R. and {Sadler}, E. and {Begam}, M.~C. and {Ianna}, P.},
        title = "{An unusual supernova in the error box of the {\ensuremath{\gamma}}-ray burst of 25 April 1998}",
      journal = {\nat},
         year = 1998,
        month = oct,
       volume = {395},
       number = {6703},
        pages = {670-672},
          doi = {10.1038/27150},
archivePrefix = {arXiv},
       eprint = {astro-ph/9806175},
 primaryClass = {astro-ph},
       adsurl = {https://ui.adsabs.harvard.edu/abs/1998Natur.395..670G}
}

@ARTICLE{2024ApJ...976...71S,
       author = {{Srinivasaragavan}, Gokul P. and {Yang}, Sheng and {Anand}, Shreya and {Sollerman}, Jesper and {Ho}, Anna Y.~Q. and {Corsi}, Alessandra and {Cenko}, S. Bradley and {Perley}, Daniel and {Schulze}, Steve and {Sanchez-Fleming}, Marquice and {Pope}, Jack and {Sarin}, Nikhil and {Omand}, Conor and {Das}, Kaustav K. and {Fremling}, Christoffer and {Andreoni}, Igor and {Bruch}, Rachel and {Burdge}, Kevin B. and {De}, Kishalay and {Gal-Yam}, Avishay and {Gangopadhyay}, Anjasha and {Graham}, Matthew J. and {Jencson}, Jacob E. and {Karambelkar}, Viraj and {Kasliwal}, Mansi M. and {Kulkarni}, S.~R. and {Martikainen}, Julia and {Sharma}, Yashvi S. and {Tzanidakis}, Anastasios and {Yan}, Lin and {Yao}, Yuhan and {Bellm}, Eric C. and {Groom}, Steven L. and {Masci}, Frank J. and {Nir}, Guy and {Purdum}, Josiah and {Smith}, Roger and {Sravan}, Niharika},
        title = "{Optical and Radio Analysis of Systematically Classified Broad-lined Type Ic Supernovae from the Zwicky Transient Facility}",
      journal = {\apj},
         year = 2024,
        month = nov,
       volume = {976},
       number = {1},
          eid = {71},
        pages = {71},
          doi = {10.3847/1538-4357/ad7fde},
archivePrefix = {arXiv},
       eprint = {2408.14586},
 primaryClass = {astro-ph.HE},
       adsurl = {https://ui.adsabs.harvard.edu/abs/2024ApJ...976...71S}
}

@ARTICLE{2025A&A...701A.128A,
       author = {{Ayala}, Bastian and {Anderson}, Joseph P. and {Pignata}, G. and {F{\"o}rster}, Francisco and {Smartt}, S.~J. and {Rest}, A. and {Solar}, Mart{\'\i}n and {Erasmus}, Nicolas and {Dastidar}, Raya and {Ramirez}, Mauricio and {Pineda-Garc{\'\i}a}, Jonathan},
        title = "{Early light curve excess in Type IIb supernovae observed with ATLAS: Qualitative constraints on progenitor systems}",
      journal = {\aap},
         year = 2025,
        month = sep,
       volume = {701},
          eid = {A128},
        pages = {A128},
          doi = {10.1051/0004-6361/202554370},
archivePrefix = {arXiv},
       eprint = {2503.05909},
 primaryClass = {astro-ph.HE},
       adsurl = {https://ui.adsabs.harvard.edu/abs/2025A&A...701A.128A}
}

@ARTICLE{2016ApJ...830...42C,
       author = {{Corsi}, A. and {Gal-Yam}, A. and {Kulkarni}, S.~R. and {Frail}, D.~A. and {Mazzali}, P.~A. and {Cenko}, S.~B. and {Kasliwal}, M.~M. and {Cao}, Y. and {Horesh}, A. and {Palliyaguru}, N. and {Perley}, D.~A. and {Laher}, R.~R. and {Taddia}, F. and {Leloudas}, G. and {Maguire}, K. and {Nugent}, P.~E. and {Sollerman}, J. and {Sullivan}, M.},
        title = "{Radio Observations of a Sample of Broad-line Type IC Supernovae Discovered by PTF/IPTF: A Search for Relativistic Explosions}",
      journal = {\apj},
         year = 2016,
        month = oct,
       volume = {830},
       number = {1},
          eid = {42},
        pages = {42},
          doi = {10.3847/0004-637X/830/1/42},
archivePrefix = {arXiv},
       eprint = {1512.01303},
 primaryClass = {astro-ph.HE},
       adsurl = {https://ui.adsabs.harvard.edu/abs/2016ApJ...830...42C}
}

@ARTICLE{2025ApJ...985..183A,
       author = {{Abac}, A.~G. and {Abbott}, R. and {Abouelfettouh}, I. and {Acernese}, F. and {Ackley}, K. and {Adhicary}, S. and {Adhikari}, N. and {Adhikari}, R.~X.},
        title = "{Search for Gravitational Waves Emitted from SN 2023ixf}",
      journal = {\apj},
         year = 2025,
        month = jun,
       volume = {985},
       number = {2},
          eid = {183},
        pages = {183},
          doi = {10.3847/1538-4357/adc681},
archivePrefix = {arXiv},
       eprint = {2410.16565},
 primaryClass = {astro-ph.HE},
       adsurl = {https://ui.adsabs.harvard.edu/abs/2025ApJ...985..183A}
}

@ARTICLE{2022ApJ...928..186A,
       author = {{Abbott}, R. and {Abbott}, T.~D. and {Acernese}, F. and {Ackley}, K. and {Adams}, C. and others},
        title = "{Search for Gravitational Waves Associated with Gamma-Ray Bursts Detected by Fermi and Swift during the LIGO-Virgo Run O3b}",
      journal = {\apj},
         year = 2022,
        month = apr,
       volume = {928},
       number = {2},
          eid = {186},
        pages = {186},
          doi = {10.3847/1538-4357/ac532b},
archivePrefix = {arXiv},
       eprint = {2111.03608},
 primaryClass = {astro-ph.HE},
       adsurl = {https://ui.adsabs.harvard.edu/abs/2022ApJ...928..186A}
}

@ARTICLE{2019ApJ...873..111I,
       author = {{Ivezi{\'c}}, {\v{Z}}eljko and {Kahn}, Steven M. and {Tyson}, J. Anthony and {Abel}, Bob and {Acosta}, Emily and {Allsman}, Robyn and {Alonso}, David and {AlSayyad}, Yusra and {Anderson}, Scott F. and {Andrew}, John},
        title = "{LSST: From Science Drivers to Reference Design and Anticipated Data Products}",
      journal = {\apj},
         year = 2019,
        month = mar,
       volume = {873},
       number = {2},
          eid = {111},
        pages = {111},
          doi = {10.3847/1538-4357/ab042c},
archivePrefix = {arXiv},
       eprint = {0805.2366},
 primaryClass = {astro-ph},
       adsurl = {https://ui.adsabs.harvard.edu/abs/2019ApJ...873..111I}
}

@ARTICLE{2024PhRvD.110d2007S,
       author = {{Szczepa{\'n}czyk}, Marek J. and {Zheng}, Yanyan and {Antelis}, Javier M. and {Benjamin}, Michael and {Bizouard}, Marie-Anne and {Casallas-Lagos}, Alejandro and {Cerd{\'a}-Dur{\'a}n}, Pablo and {Davis}, Derek and {Gondek-Rosi{\'n}ska}, Dorota and {Klimenko}, Sergey and {Moreno}, Claudia and {Obergaulinger}, Martin and {Powell}, Jade and {Ramirez}, Dymetris and {Ratto}, Brad and {Richardson}, Colter and {Rijal}, Abhinav and {Stuver}, Amber L. and {Szewczyk}, Pawe{\l} and {Vedovato}, Gabriele and {Zanolin}, Michele and {Bartos}, Imre and {Bhaumik}, Shubhagata and {Bulik}, Tomasz and {Drago}, Marco and {Font}, Jos{\'e} A. and {De Colle}, Fabio and {Garc{\'\i}a-Bellido}, Juan and {Gayathri}, V. and {Hughey}, Brennan and {Mitselmakher}, Guenakh and {Mishra}, Tanmaya and {Mukherjee}, Soma and {Nguyen}, Quynh Lan and {Chan}, Man Leong and {Di Palma}, Irene and {Piotrzkowski}, Brandon J. and {Singh}, Neha},
        title = "{Optically targeted search for gravitational waves emitted by core-collapse supernovae during the third observing run of Advanced LIGO and Advanced Virgo}",
      journal = {\prd},
         year = 2024,
        month = aug,
       volume = {110},
       number = {4},
          eid = {042007},
        pages = {042007},
          doi = {10.1103/PhysRevD.110.042007},
archivePrefix = {arXiv},
       eprint = {2305.16146},
 primaryClass = {astro-ph.HE},
       adsurl = {https://ui.adsabs.harvard.edu/abs/2024PhRvD.110d2007S}
}

@article{Pratten:2020ceb,
 archiveprefix = {arXiv},
 author = {Pratten, Geraint and others},
 doi = {10.1103/PhysRevD.103.104056},
 eprint = {2004.06503},
 journal = {Phys. Rev. D},
 number = {10},
 pages = {104056},
 primaryclass = {gr-qc},
 title = {{Computationally efficient models for the dominant and subdominant harmonic modes of precessing binary black holes}},
 volume = {103},
 year = {2021}
}

@article{Husa:2015iqa,
 archiveprefix = {arXiv},
 author = {Husa, Sascha and Khan, Sebastian and Hannam, Mark and P\"urrer, Michael and Ohme, Frank and Jim\'enez Forteza, Xisco and Boh\'e, Alejandro},
 doi = {10.1103/PhysRevD.93.044006},
 eprint = {1508.07250},
 journal = {Phys. Rev. D},
 number = {4},
 pages = {044006},
 primaryclass = {gr-qc},
 title = {{Frequency-domain gravitational waves from nonprecessing black-hole binaries. I. New numerical waveforms and anatomy of the signal}},
 volume = {93},
 year = {2016}
}

@article{Khan:2015jqa,
 archiveprefix = {arXiv},
 author = {Khan, Sebastian and Husa, Sascha and Hannam, Mark and Ohme, Frank and P\"urrer, Michael and Jim\'enez Forteza, Xisco and Boh\'e, Alejandro},
 doi = {10.1103/PhysRevD.93.044007},
 eprint = {1508.07253},
 journal = {Phys. Rev. D},
 number = {4},
 pages = {044007},
 primaryclass = {gr-qc},
 title = {{Frequency-domain gravitational waves from nonprecessing black-hole binaries. II. A phenomenological model for the advanced detector era}},
 volume = {93},
 year = {2016}
}

@article{Fairhurst:2009tc,
    author = "Fairhurst, Stephen",
    title = "{Triangulation of gravitational wave sources with a network of detectors}",
    eprint = "0908.2356",
    archivePrefix = "arXiv",
    primaryClass = "gr-qc",
    doi = "10.1088/1367-2630/11/12/123006",
    journal = "New J. Phys.",
    volume = "11",
    pages = "123006",
    year = "2009",
    note = "[Erratum: New J.Phys. 13, 069602 (2011)]"
}

@ARTICLE{SunH+15,
       author = {{Sun}, Hui and {Zhang}, Bing and {Li}, Zhuo},
        title = "{Extragalactic High-energy Transients: Event Rate Densities and Luminosity Functions}",
      journal = {\apj},
         year = 2015,
        month = oct,
       volume = {812},
       number = {1},
          eid = {33},
        pages = {33},
          doi = {10.1088/0004-637X/812/1/33},
archivePrefix = {arXiv},
       eprint = {1509.01592},
 primaryClass = {astro-ph.HE},
       adsurl = {https://ui.adsabs.harvard.edu/abs/2015ApJ...812...33S}
}

@ARTICLE{2005PhRvD..72d2002A,
       author = {{LIGO Scientific Collaboration}},
        title = "{Search for gravitational waves associated with the gamma ray burst GRB030329 using the LIGO detectors}",
      journal = {\prd},
         year = 2005,
        month = aug,
       volume = {72},
       number = {4},
          eid = {042002},
        pages = {042002},
          doi = {10.1103/PhysRevD.72.042002},
archivePrefix = {arXiv},
       eprint = {gr-qc/0501068},
 primaryClass = {gr-qc},
       adsurl = {https://ui.adsabs.harvard.edu/abs/2005PhRvD..72d2002A}
}

@ARTICLE{LloydRonningN+19,
       author = {{Lloyd-Ronning}, Nicole M. and {Aykutalp}, Aycin and {Johnson}, Jarrett L.},
        title = "{On the cosmological evolution of long gamma-ray burst properties}",
      journal = {\mnras},
         year = 2019,
        month = oct,
       volume = {488},
       number = {4},
        pages = {5823-5832},
          doi = {10.1093/mnras/stz2155},
archivePrefix = {arXiv},
       eprint = {1906.02278},
 primaryClass = {astro-ph.HE},
       adsurl = {https://ui.adsabs.harvard.edu/abs/2019MNRAS.488.5823L}
}

@ARTICLE{RuffiniR+18,
       author = {{Ruffini}, R. and {Rodriguez}, J. and {Muccino}, M. and {Rueda}, J.~A. and {Aimuratov}, Y. and {Barres de Almeida}, U. and {Becerra}, L. and {Bianco}, C.~L. and {Cherubini}, C. and {Filippi}, S. and {Gizzi}, D. and {Kovacevic}, M. and {Moradi}, R. and {Oliveira}, F.~G. and {Pisani}, G.~B. and {Wang}, Y.},
        title = "{On the Rate and on the Gravitational Wave Emission of Short and Long GRBs}",
      journal = {\apj},
         year = 2018,
        month = may,
       volume = {859},
       number = {1},
          eid = {30},
        pages = {30},
          doi = {10.3847/1538-4357/aabee4},
archivePrefix = {arXiv},
       eprint = {1602.03545},
 primaryClass = {astro-ph.HE},
       adsurl = {https://ui.adsabs.harvard.edu/abs/2018ApJ...859...30R}
}

@ARTICLE{LiangE+07,
       author = {{Liang}, Enwei and {Zhang}, Bing and {Virgili}, Francisco and {Dai}, Z.~G.},
        title = "{Low-Luminosity Gamma-Ray Bursts as a Unique Population: Luminosity Function, Local Rate, and Beaming Factor}",
      journal = {\apj},
         year = 2007,
        month = jun,
       volume = {662},
       number = {2},
        pages = {1111-1118},
          doi = {10.1086/517959},
archivePrefix = {arXiv},
       eprint = {astro-ph/0605200},
 primaryClass = {astro-ph},
       adsurl = {https://ui.adsabs.harvard.edu/abs/2007ApJ...662.1111L}
}

@ARTICLE{PescalliA+16,
       author = {{Pescalli}, A. and {Ghirlanda}, G. and {Salvaterra}, R. and {Ghisellini}, G. and {Vergani}, S.~D. and {Nappo}, F. and {Salafia}, O.~S. and {Melandri}, A. and {Covino}, S. and {G{\"o}tz}, D.},
        title = "{The rate and luminosity function of long gamma ray bursts}",
      journal = {\aap},
         year = 2016,
        month = mar,
       volume = {587},
          eid = {A40},
        pages = {A40},
          doi = {10.1051/0004-6361/201526760},
archivePrefix = {arXiv},
       eprint = {1506.05463},
 primaryClass = {astro-ph.HE},
       adsurl = {https://ui.adsabs.harvard.edu/abs/2016A&A...587A..40P}
}

@ARTICLE{ODwyerT+26,
       author = {{O'Dwyer}, Tanner and {Corsi}, Alessandra and {Yang}, Sheng and {Anand}, Shreya and {Cenko}, S. Bradley and {Srinivasaragavan}, Gokul P. and {Ho}, Anna Y.~Q. and {Sollerman}, Jesper and {Zhou}, Bei and {Balasubramanian}, Arvind and {Chang}, Po-Wen and {Kamionkowski}, Marc and {Perley}, Daniel and {Laher}, Russ R. and {Murase}, Kohta and {Masci}, Frank J. and {Kasliwal}, Mansi M. and {Purdum}, Josiah N. and {Graham}, Matthew J.},
        title = "{A Search for Successful and Choked Jets in Nearby Broad-lined Type Ic Supernovae}",
      journal = {\apj},
         year = 2026,
        month = may,
       volume = {1002},
       number = {2},
          eid = {194},
        pages = {194},
          doi = {10.3847/1538-4357/ae522d},
archivePrefix = {arXiv},
       eprint = {2512.08822},
 primaryClass = {astro-ph.HE},
       adsurl = {https://ui.adsabs.harvard.edu/abs/2026ApJ..1002..194O}
}

@ARTICLE{DongX+23,
       author = {{Dong}, X.~F. and {Zhang}, Z.~B. and {Li}, Q.~M. and {Huang}, Y.~F. and {Bian}, K.},
        title = "{The Origin of Low-redshift Event Rate Excess as Revealed by the Low-luminosity Gamma-Ray Bursts}",
      journal = {\apj},
         year = 2023,
        month = nov,
       volume = {958},
       number = {1},
          eid = {37},
        pages = {37},
          doi = {10.3847/1538-4357/acf852},
archivePrefix = {arXiv},
       eprint = {2305.11380},
 primaryClass = {astro-ph.HE},
       adsurl = {https://ui.adsabs.harvard.edu/abs/2023ApJ...958...37D}
}

@ARTICLE{MarguttiR+14,
       author = {{Margutti}, R. and {Milisavljevic}, D. and {Soderberg}, A.~M. and {Guidorzi}, C. and {Morsony}, B.~J. and {Sanders}, N. and {Chakraborti}, S. and {Ray}, A. and {Kamble}, A. and {Drout}, M. and {Parrent}, J. and {Zauderer}, A. and {Chomiuk}, L.},
        title = "{Relativistic Supernovae have Shorter-lived Central Engines or More Extended Progenitors: The Case of SN 2012ap}",
      journal = {\apj},
         year = 2014,
        month = dec,
       volume = {797},
       number = {2},
          eid = {107},
        pages = {107},
          doi = {10.1088/0004-637X/797/2/107},
archivePrefix = {arXiv},
       eprint = {1402.6344},
 primaryClass = {astro-ph.HE},
       adsurl = {https://ui.adsabs.harvard.edu/abs/2014ApJ...797..107M}
}

@ARTICLE{JapeljJ+18,
       author = {{Japelj}, J. and {Vergani}, S.~D. and {Salvaterra}, R. and {Renzo}, M. and {Zapartas}, E. and {de Mink}, S.~E. and {Kaper}, L. and {Zibetti}, S.},
        title = "{Host galaxies of SNe Ic-BL with and without long gamma-ray bursts}",
      journal = {\aap},
         year = 2018,
        month = sep,
       volume = {617},
          eid = {A105},
        pages = {A105},
          doi = {10.1051/0004-6361/201833209},
archivePrefix = {arXiv},
       eprint = {1806.10613},
 primaryClass = {astro-ph.HE},
       adsurl = {https://ui.adsabs.harvard.edu/abs/2018A&A...617A.105J}
}

@ARTICLE{CamisascaA+23,
       author = {{Camisasca}, A.~E. and {Guidorzi}, C. and {Amati}, L. and {Frontera}, F. and {Song}, X.~Y. and {Xiao}, S. and {Xiong}, S.~L. and {Zhang}, S.~N. and {Margutti}, R. and {Kobayashi}, S. and {Mundell}, C.~G. and {Ge}, M.~Y. and {Gomboc}, A. and {Jia}, S.~M. and {Jordana-Mitjans}, N. and {Li}, C.~K. and {Li}, X.~B. and {Maccary}, R. and {Shrestha}, M. and {Xue}, W.~C. and {Zhang}, S.},
        title = "{GRB minimum variability timescale with Insight-HXMT and Swift. Implications for progenitor models, dissipation physics, and GRB classifications}",
      journal = {\aap},
         year = 2023,
        month = mar,
       volume = {671},
          eid = {A112},
        pages = {A112},
          doi = {10.1051/0004-6361/202245657},
archivePrefix = {arXiv},
       eprint = {2301.01176},
 primaryClass = {astro-ph.HE},
       adsurl = {https://ui.adsabs.harvard.edu/abs/2023A&A...671A.112C}
}

@ARTICLE{CorsiA+23,
       author = {{Corsi}, Alessandra and {Ho}, Anna Y.~Q. and {Cenko}, S. Bradley and {Kulkarni}, Shrinivas R. and {Anand}, Shreya and {Yang}, Sheng and {Sollerman}, Jesper and {Srinivasaragavan}, Gokul P. and {Omand}, Conor M.~B. and {Balasubramanian}, Arvind and {Frail}, Dale A. and {Fremling}, Christoffer and {Perley}, Daniel A. and {Yao}, Yuhan and {Dahiwale}, Aishwarya S. and {De}, Kishalay and {Dugas}, Alison and {Hankins}, Matthew and {Jencson}, Jacob and {Kasliwal}, Mansi M. and {Tzanidakis}, Anastasios and {Bellm}, Eric C. and {Laher}, Russ R. and {Masci}, Frank J. and {Purdum}, Josiah N. and {Regnault}, Nicolas},
        title = "{A Search for Relativistic Ejecta in a Sample of ZTF Broad-lined Type Ic Supernovae}",
      journal = {\apj},
         year = 2023,
        month = aug,
       volume = {953},
       number = {2},
          eid = {179},
        pages = {179},
          doi = {10.3847/1538-4357/acd3f2},
archivePrefix = {arXiv},
       eprint = {2210.09536},
 primaryClass = {astro-ph.HE},
       adsurl = {https://ui.adsabs.harvard.edu/abs/2023ApJ...953..179C}
}

@ARTICLE{2025ApJ...994L..45F,
       author = {{Franz}, Noah and {Subrayan}, Bhagya and {Kilpatrick}, Charles D. and {Hosseinzadeh}, Griffin and {Sand}, David J. and {Alexander}, Kate D. and {Fong}, Wen-fai and {Christy}, Collin T. and {Pearson}, Jeniveve and {Laskar}, Tanmoy and {Hsu}, Brian and {Rastinejad}, Jillian and {Lundquist}, Michael J. and {Berger}, Edo and {Bostroem}, K. Azalee and {Bom}, Clecio R. and {Darc}, Phelipe and {Gurwell}, Mark and {Schimpf}, Shelbi Hostler and {Keating}, Garrett K. and {Noel}, Phillip and {Ransome}, Conor and {Rao}, Ramprasad and {Santana-Silva}, Luidhy and {Santos}, A. Souza and {Shrestha}, Manisha and {Anche}, Ramya and {Andrews}, Jennifer E. and {Borthakur}, Sanchayeeta and {Butler}, Nathaniel R. and {Coppejans}, Deanne L. and {Daly}, Philip N. and {Daniel}, Kathryne J. and {Duffell}, Paul C. and {Eftekhari}, Tarraneh and {Fields}, Carl E. and {Gagliano}, Alexander T. and {Golay}, Walter W. and {Grichener}, Aldana and {Hamden}, Erika T. and {Hiramatsu}, Daichi and {Kumar}, Harsh and {Manikantan}, Vikram and {Margutti}, Raffaella and {Paschalidis}, Vasileios and {Paterson}, Kerry and {Reichart}, Daniel E. and {Renzo}, Mathieu and {Salmas}, Kali and {Schroeder}, Genevieve and {Smith}, Nathan and {Spekkens}, Kristine and {Strader}, Jay and {Trilling}, David E. and {Vieira}, Nicholas and {Weiner}, Benjamin and {Williams}, Peter K.~G.},
        title = "{Optimizing Kilonova Searches: A Case Study of the Type IIb SN 2025ulz in the Localization Volume of the Low-significance Gravitational Wave Event S250818k}",
      journal = {\apjl},
         year = 2025,
        month = dec,
       volume = {994},
       number = {2},
          eid = {L45},
        pages = {L45},
          doi = {10.3847/2041-8213/ae17a8},
archivePrefix = {arXiv},
       eprint = {2510.17104},
 primaryClass = {astro-ph.HE},
       adsurl = {https://ui.adsabs.harvard.edu/abs/2025ApJ...994L..45F}
}

@ARTICLE{2008PhRvD..77h2001D,
       author = {{Dhurandhar}, Sanjeev and {Krishnan}, Badri and {Mukhopadhyay}, Himan and {Whelan}, John T.},
        title = "{Cross-correlation search for periodic gravitational waves}",
      journal = {\prd},
         year = 2008,
        month = apr,
       volume = {77},
       number = {8},
          eid = {082001},
        pages = {082001},
          doi = {10.1103/PhysRevD.77.082001},
archivePrefix = {arXiv},
       eprint = {0712.1578},
 primaryClass = {gr-qc},
       adsurl = {https://ui.adsabs.harvard.edu/abs/2008PhRvD..77h2001D}
}

@ARTICLE{2019PhRvD.100l4041S,
       author = {{Sowell}, Eric and {Corsi}, Alessandra and {Coyne}, Robert},
        title = "{Multiwaveform cross-correlation search method for intermediate-duration gravitational waves from gamma-ray bursts}",
      journal = {\prd},
         year = 2019,
        month = dec,
       volume = {100},
       number = {12},
          eid = {124041},
        pages = {124041},
          doi = {10.1103/PhysRevD.100.124041},
archivePrefix = {arXiv},
       eprint = {1906.03998},
 primaryClass = {astro-ph.HE},
       adsurl = {https://ui.adsabs.harvard.edu/abs/2019PhRvD.100l4041S}
}

@ARTICLE{2026PhRvD.113j3034K,
       author = {{Khanam}, Tanazza and {Corsi}, Alessandra and {Coyne}, Robert and {Pierre}, Michael St.},
        title = "{Searches for postmerger gravitational waves with CoCoA: Sensitivity projections across large template banks for current and next-generation detectors}",
      journal = {\prd},
         year = 2026,
        month = may,
       volume = {113},
       number = {10},
          eid = {103034},
        pages = {103034},
          doi = {10.1103/yc6h-g481},
archivePrefix = {arXiv},
       eprint = {2511.21941},
 primaryClass = {gr-qc},
       adsurl = {https://ui.adsabs.harvard.edu/abs/2026PhRvD.113j3034K}
}

@ARTICLE{2025PhRvD.112j2005A,
       author = {{Ligo Scientific Collaboration} and {VIRGO Collaboration} and {Kagra Collaboration}},
        title = "{All-sky search for short gravitational-wave bursts in the first part of the fourth LIGO-Virgo-KAGRA observing run}",
      journal = {\prd},
         year = 2025,
        month = nov,
       volume = {112},
       number = {10},
          eid = {102005},
        pages = {102005},
          doi = {10.1103/wjdz-jdby},
archivePrefix = {arXiv},
       eprint = {2507.12374},
 primaryClass = {astro-ph.HE},
       adsurl = {https://ui.adsabs.harvard.edu/abs/2025PhRvD.112j2005A}
}

@ARTICLE{2026arXiv260405128O,
       author = {{O'Dwyer}, Tanner and {Corsi}, Alessandra and {Yadav}, Deepika and {Mooley}, Kunal P. and {Baer-Way}, Raphael and {Chandra}, Poonam and {Hallinan}, Gregg and {Kasliwal}, Mansi M. and {Rhodes}, Lauren and {Smirnov}, Oleg M. and {Lazzati}, Davide and {van Leeuwen}, Joeri and {Deller}, Adam and {Atri}, Pikky and {Khanam}, Tanazza},
        title = "{Identification of a Radio Counterpart to SN 2025ulz in the S250818k Localization Area}",
      journal = {arXiv e-prints},
         year = 2026,
        month = apr,
          eid = {arXiv:2604.05128},
        pages = {arXiv:2604.05128},
          doi = {10.48550/arXiv.2604.05128},
archivePrefix = {arXiv},
       eprint = {2604.05128},
 primaryClass = {astro-ph.HE},
       adsurl = {https://ui.adsabs.harvard.edu/abs/2026arXiv260405128O}
}

@ARTICLE{2025GCN.41437....1L,
       author = {{Ligo Scientific Collaboration} and {VIRGO Collaboration} and {Kagra Collaboration}},
        title = "{LIGO/Virgo/KAGRA S250818k: Properties of the low-significance GW compact binary merger candidate potentially associated with AT 2025ulz}",
      journal = {GRB Coordinates Network},
         year = 2025,
        month = aug,
       volume = {41437},
        pages = {1},
       adsurl = {https://ui.adsabs.harvard.edu/abs/2025GCN.41437....1L}
}

@ARTICLE{2025GCN.41414....1S,
       author = {{Stein}, Robert and {Ahumada}, Tom{\'a}s and {Kasliwal}, Mansi and {Du Laz}, Theophile and {Pathak}, Utkarsh and {Swain}, Vishwajeet and {Salgundi}, Anirudh and {Bhalerao}, Varun and {Hall}, Xander J. and {Ztf Collaboration} and {Growth Collaboration}},
        title = "{LIGO/Virgo/KAGRA S250818k: Candidates from the Zwicky Transient Facility}",
      journal = {GRB Coordinates Network},
         year = 2025,
        month = aug,
       volume = {41414},
        pages = {1},
       adsurl = {https://ui.adsabs.harvard.edu/abs/2025GCN.41414....1S}
}

@ARTICLE{2015CQGra..32g4001L,
       author = {{LIGO Scientific Collaboration} },
        title = "{Advanced LIGO}",
      journal = {Classical and Quantum Gravity},
         year = 2015,
        month = apr,
       volume = {32},
       number = {7},
          eid = {074001},
        pages = {074001},
          doi = {10.1088/0264-9381/32/7/074001},
archivePrefix = {arXiv},
       eprint = {1411.4547},
 primaryClass = {gr-qc},
       adsurl = {https://ui.adsabs.harvard.edu/abs/2015CQGra..32g4001L}
}

@misc{lalsuite,
    author       = "{LIGO Scientific Collaboration} and {Virgo Collaboration} and {KAGRA Collaboration}",
    title        = "{LVK} {A}lgorithm {L}ibrary - {LALS}uite",
    howpublished = "Free software (GPL)",
    doi          = "10.7935/GT1W-FZ16",
    year         = "2018"
  }

@ARTICLE{2015CQGra..32b4001A,
       author = {{Acernese}, F. and {Agathos}, M. and {Agatsuma}, K. and {Aisa}, D. and {Allemandou}, N. and others},
        title = "{Advanced Virgo: a second-generation interferometric gravitational wave detector}",
      journal = {Classical and Quantum Gravity},
         year = 2015,
        month = jan,
       volume = {32},
       number = {2},
          eid = {024001},
        pages = {024001},
          doi = {10.1088/0264-9381/32/2/024001},
archivePrefix = {arXiv},
       eprint = {1408.3978},
 primaryClass = {gr-qc},
       adsurl = {https://ui.adsabs.harvard.edu/abs/2015CQGra..32b4001A}
}

@ARTICLE{2021PTEP.2021eA101A,
       author = {{Akutsu}, T. and {Ando}, M. and {Arai}, K. and {Arai}, Y. and {Araki}, S. and {Araya}, A. and {Aritomi}},
        title = "{Overview of KAGRA: Detector design and construction history}",
      journal = {Progress of Theoretical and Experimental Physics},
         year = 2021,
        month = may,
       volume = {2021},
       number = {5},
          eid = {05A101},
        pages = {05A101},
          doi = {10.1093/ptep/ptaa125},
archivePrefix = {arXiv},
       eprint = {2005.05574},
 primaryClass = {physics.ins-det},
       adsurl = {https://ui.adsabs.harvard.edu/abs/2021PTEP.2021eA101A}
}

@ARTICLE{2017ApJ...848L..12A,
       author = {{The LIGO Scientific Collaboration} and {the Virgo Collaboration} and others},
        title = "{Multi-messenger Observations of a Binary Neutron Star Merger}",
      journal = {\apjl},
         year = 2017,
        month = oct,
       volume = {848},
       number = {2},
          eid = {L12},
        pages = {L12},
          doi = {10.3847/2041-8213/aa91c9},
archivePrefix = {arXiv},
       eprint = {1710.05833},
 primaryClass = {astro-ph.HE},
       adsurl = {https://ui.adsabs.harvard.edu/abs/2017ApJ...848L..12A}
}

@ARTICLE{2017ApJ...848L..13A,
       author = {{The LIGO Scientific Collaboration} and {the Virgo Collaboration} and others},
        title = "{Gravitational Waves and Gamma-Rays from a Binary Neutron Star Merger: GW170817 and GRB 170817A}",
      journal = {\apjl},
         year = 2017,
        month = oct,
       volume = {848},
       number = {2},
          eid = {L13},
        pages = {L13},
          doi = {10.3847/2041-8213/aa920c},
archivePrefix = {arXiv},
       eprint = {1710.05834},
 primaryClass = {astro-ph.HE},
       adsurl = {https://ui.adsabs.harvard.edu/abs/2017ApJ...848L..13A}
}

@ARTICLE{2017PhRvL.119p1101A,
       author = {{The LIGO Scientific Collaboration} and {the Virgo Collaboration}},
        title = "{GW170817: Observation of Gravitational Waves from a Binary Neutron Star Inspiral}",
      journal = {\prl},
         year = 2017,
        month = oct,
       volume = {119},
       number = {16},
          eid = {161101},
        pages = {161101},
          doi = {10.1103/PhysRevLett.119.161101},
archivePrefix = {arXiv},
       eprint = {1710.05832},
 primaryClass = {gr-qc},
       adsurl = {https://ui.adsabs.harvard.edu/abs/2017PhRvL.119p1101A}
}

@ARTICLE{2025arXiv250818083T,
       author = {{The LIGO Scientific Collaboration} and {the Virgo Collaboration} and {the KAGRA Collaboration} },
        title = "{GWTC-4.0: Population Properties of Merging Compact Binaries}",
      journal = {arXiv e-prints},
         year = 2025,
        month = aug,
          eid = {arXiv:2508.18083},
        pages = {arXiv:2508.18083},
          doi = {10.48550/arXiv.2508.18083},
archivePrefix = {arXiv},
       eprint = {2508.18083},
 primaryClass = {astro-ph.HE},
       adsurl = {https://ui.adsabs.harvard.edu/abs/2025arXiv250818083T}
}

@article{Fairhurst:2017mvj,
    author = "Fairhurst, Stephen",
    title = "{Localization of transient gravitational wave sources:
    beyond triangulation}",
    eprint = "1712.04724",
    archivePrefix = "arXiv",
    primaryClass = "gr-qc",
    reportNumber = "LIGO-P1300174",
    doi = "10.1088/1361-6382/aab675",
    journal = "Class. Quant. Grav.",
    volume = "35",
    number = "10",
    pages = "105002",
    year = "2018"
}

@article{Fairhurst:2023idl,
    author = "Fairhurst, Stephen and Hoy, Charlie and Green, Rhys
    and Mills, Cameron and Usman, Samantha A.",
    title = "{Simple parameter estimation using observable features of
    gravitational-wave signals}",
    eprint = "2304.03731",
    archivePrefix = "arXiv",
    primaryClass = "gr-qc",
    doi = "10.1103/PhysRevD.108.082006",
    journal = "Phys. Rev. D",
    volume = "108",
    number = "8",
    pages = "082006",
    year = "2023"
}

@article{ET:2025xjr,
    author = "Abac, Adrian and others",
    title = "{The Science of the Einstein Telescope}",
      journal = {\jcap},
         year = 2026,
        month = mar,
       volume = {2026},
       number = {3},
          eid = {081},
        pages = {081},
          doi = {10.1088/1475-7516/2026/03/081},
archivePrefix = {arXiv},
       eprint = {2503.12263},
 primaryClass = {gr-qc},
       adsurl = {https://ui.adsabs.harvard.edu/abs/2026JCAP...03..081A}
}

@article{Allen:2005fk,
    author = "Allen, Bruce and Anderson, Warren G. and Brady, Patrick R. and Brown, Duncan A. and Creighton, Jolien D. E.",
    title = "{FINDCHIRP: An Algorithm for detection of gravitational waves from inspiraling compact binaries}",
    eprint = "gr-qc/0509116",
    archivePrefix = "arXiv",
    doi = "10.1103/PhysRevD.85.122006",
    journal = "Phys. Rev. D",
    volume = "85",
    pages = "122006",
    year = "2012"
}

@misc{O4c_sensitivity,
    author = {{{LIGO Scientific Collaboration}}},
    title = {O4c Reference Sensitivity},
    url = {https://dcc.ligo.org/LIGO-T2500364},
    year = {2025}
}

@misc{ASharp_sensitivity,
 author = {{{LIGO Scientific Collaboration}}},
 title = {A\# Strain Sensitivity},
 url = {https://dcc.ligo.org/LIGO-T2300041/public},
 year = {2026}
}

@misc{O5c_sensitivity,
    author = {{{LIGO Scientific Collaboration}}},
    title = {A+/O5 strain curve projections},
    url = {https://dcc.ligo.org/LIGO-T2500310/public},
    year = {2025}
}

@misc{CE_sensitivity,
    author = {{{CE Consortium}}},
    title = {Cosmic Explorer Strain Sensitivity},
    url = {https://dcc.cosmicexplorer.org/CE-T2000017-v8/public},
    year = {2024}
}

@misc{ET_sensitivity_CoBA,
    author = {{{ET Collaboration}}},
    title = {ET sensitivity curves used for CoBA Science Study},
    url = {https://apps.et-gw.eu/tds/?r=18213},
    year = {2023}
}

@ARTICLE{LIGO+25_Rates,
       author = {{Abac}, A.~G. and {Abouelfettouh}, I. and {Acernese}, F. and {Ackley}, K. and {Adamcewicz}, C. and {Adhicary}, S. and {Adhikari}, D. and {Adhikari}, N. and {Adhikari}, R.~X.},
        title = "{GWTC-4.0: Updating the Gravitational-Wave Transient Catalog with Observations from the First Part of the Fourth LIGO-Virgo-KAGRA Observing Run}",
      journal = {arXiv e-prints},
         year = 2025,
        month = aug,
          eid = {arXiv:2508.18082},
        pages = {arXiv:2508.18082},
          doi = {10.48550/arXiv.2508.18082},
archivePrefix = {arXiv},
       eprint = {2508.18082},
 primaryClass = {gr-qc},
       adsurl = {https://ui.adsabs.harvard.edu/abs/2025arXiv250818082T}
}

@ARTICLE{LIGO+23_Rates,
       author = {{Abbott}, R. and {Abbott}, T.~D. and {Acernese}, F. and {Ackley}, K. and {Adams}, C. and {Adhikari}, N. and {Adhikari}, R.~X. and {Adya}, V.~B. and {Affeldt}, C. and {Agarwal}, D. and {Agathos}, M. and {Agatsuma}, K.},
        title = "{Population of Merging Compact Binaries Inferred Using Gravitational Waves through GWTC-3}",
      journal = {Physical Review X},
         year = 2023,
        month = jan,
       volume = {13},
       number = {1},
          eid = {011048},
        pages = {011048},
          doi = {10.1103/PhysRevX.13.011048},
archivePrefix = {arXiv},
       eprint = {2111.03634},
 primaryClass = {astro-ph.HE},
       adsurl = {https://ui.adsabs.harvard.edu/abs/2023PhRvX..13a1048A}
}

@ARTICLE{ABbott+22_O3,
       author = {{Abbott}, R. and {Abe}, H. and {Acernese}, F. and {Ackley}, K. and {Adhikari}, N. and {Adhikari}, R.~X. and {Adkins}, V.~K. and {Adya}, V.~B. and {Affeldt}, C. and {Agarwal}, D. and {Agathos}, M.},
        title = "{All-sky search for continuous gravitational waves from isolated neutron stars using Advanced LIGO and Advanced Virgo O3 data}",
      journal = {\prd},
         year = 2022,
        month = nov,
       volume = {106},
       number = {10},
          eid = {102008},
        pages = {102008},
          doi = {10.1103/PhysRevD.106.102008},
archivePrefix = {arXiv},
       eprint = {2201.00697},
 primaryClass = {gr-qc},
       adsurl = {https://ui.adsabs.harvard.edu/abs/2022PhRvD.106j2008A}
}

@ARTICLE{Abbott+19_LongLivedBNS,
       author = {{Abbott}, B.~P. and {Abbott}, R. and {Abbott}, T.~D. and {Acernese}, F. and {Ackley}, K. and {Adams}, C. and {Adams}, T. and {Addesso}, P. and {Adhikari}, R.~X. and {Adya}, V.~B.},
        title = "{Search for Gravitational Waves from a Long-lived Remnant of the Binary Neutron Star Merger GW170817}",
      journal = {\apj},
         year = 2019,
        month = apr,
       volume = {875},
       number = {2},
          eid = {160},
        pages = {160},
          doi = {10.3847/1538-4357/ab0f3d},
archivePrefix = {arXiv},
       eprint = {1810.02581},
 primaryClass = {gr-qc},
       adsurl = {https://ui.adsabs.harvard.edu/abs/2019ApJ...875..160A}
}

@ARTICLE{NaozS2016,
       author = {{Naoz}, Smadar},
        title = "{The Eccentric Kozai-Lidov Effect and Its Applications}",
      journal = {\araa},
         year = 2016,
        month = sep,
       volume = {54},
        pages = {441-489},
          doi = {10.1146/annurev-astro-081915-023315},
archivePrefix = {arXiv},
       eprint = {1601.07175},
 primaryClass = {astro-ph.EP},
       adsurl = {https://ui.adsabs.harvard.edu/abs/2016ARA&A..54..441N}
}

@ARTICLE{BaiottiL_RezzollaL17,
       author = {{Baiotti}, Luca and {Rezzolla}, Luciano},
        title = "{Binary neutron star mergers: a review of Einstein{\textquoteright}s richest laboratory}",
      journal = {Reports on Progress in Physics},
         year = 2017,
        month = sep,
       volume = {80},
       number = {9},
          eid = {096901},
        pages = {096901},
          doi = {10.1088/1361-6633/aa67bb},
archivePrefix = {arXiv},
       eprint = {1607.03540},
 primaryClass = {gr-qc},
       adsurl = {https://ui.adsabs.harvard.edu/abs/2017RPPh...80i6901B}
}

@ARTICLE{RodriguezL+18,
       author = {{Rodriguez}, Carl L. and {Amaro-Seoane}, Pau and {Chatterjee}, Sourav and {Kremer}, Kyle and {Rasio}, Frederic A. and {Samsing}, Johan and {Ye}, Claire S. and {Zevin}, Michael},
        title = "{Post-Newtonian dynamics in dense star clusters: Formation, masses, and merger rates of highly-eccentric black hole binaries}",
      journal = {\prd},
         year = 2018,
        month = dec,
       volume = {98},
       number = {12},
          eid = {123005},
        pages = {123005},
          doi = {10.1103/PhysRevD.98.123005},
archivePrefix = {arXiv},
       eprint = {1811.04926},
 primaryClass = {astro-ph.HE},
       adsurl = {https://ui.adsabs.harvard.edu/abs/2018PhRvD..98l3005R}
}

@ARTICLE{AntogniniJ+14,
       author = {{Antognini}, Joe M. and {Shappee}, Benjamin J. and {Thompson}, Todd A. and {Amaro-Seoane}, Pau},
        title = "{Rapid eccentricity oscillations and the mergers of compact objects in hierarchical triples}",
      journal = {\mnras},
         year = 2014,
        month = mar,
       volume = {439},
       number = {1},
        pages = {1079-1091},
          doi = {10.1093/mnras/stu039},
archivePrefix = {arXiv},
       eprint = {1308.5682},
 primaryClass = {astro-ph.HE},
       adsurl = {https://ui.adsabs.harvard.edu/abs/2014MNRAS.439.1079A}
}

@ARTICLE{KasliwalM+25_S250818k,
       author = {{Kasliwal}, Mansi M. and {Ahumada}, Tomas and {Stein}, Robert and {Karambelkar}, Viraj and {Hall}, Xander J. and {Singh}, Avinash and {Fremling}, Christoffer and {Metzger}, Brian D. and {Bulla}, Mattia and {Swain}, Vishwajeet and {Antier}, Sarah and {Pillas}, Marion and {Busmann}, Malte and {Freeburn}, James and {Karpov}, Sergey and {Bochenek}, Aleksandra and {O'Connor}, Brendan and {Perley}, Daniel A. and {Akl}, Dalya and {Anand}, Shreya and {Toivonen}, Andrew and {Rose}, Sam and {Jegou du Laz}, Theophile and {Liu}, Chang and {Das}, Kaustav and {Sharma Chaudhary}, Sushant and {Barna}, Tyler and {Pawan Saikia}, Aditya and {Andreoni}, Igor and {Bellm}, Eric C. and {Bhalerao}, Varun and {Cenko}, S. Bradley and {Coughlin}, Michael W. and {Gruen}, Daniel and {Kasen}, Daniel and {Miller}, Adam A. and {Nissanke}, Samaya and {Palmese}, Antonella and {Sollerman}, Jesper and {Sravan}, Niharika and {Anupama}, G.~C. and {Banerjee}, Smaranika and {Barway}, Sudhanshu and {Bloom}, Joshua S. and {Cabrera}, Tomas and {Chen}, Tracy and {Copperwheat}, Chris and {Corsi}, Alessandra and {Dekany}, Richard and {Earley}, Nicholas and {Graham}, Matthew and {Hello}, Patrice and {Helou}, George and {Hu}, Lei and {Kini}, Yves and {Mahabal}, Ashish and {Masci}, Frank and {Mohan}, Tanishk and {Pletskova}, Natalya and {Purdum}, Josiah and {Qin}, Yu-Jing and {Rehemtulla}, Nabeel and {Salgundi}, Anirudh and {Wang}, Yuankun},
        title = "{ZTF25abjmnps (AT2025ulz) and S250818k: A Candidate Superkilonova from a Sub-threshold Sub-Solar Gravitational Wave Trigger}",
      journal = {arXiv e-prints},
         year = 2025,
        month = oct,
          eid = {arXiv:2510.23732},
        pages = {arXiv:2510.23732},
          doi = {10.48550/arXiv.2510.23732},
archivePrefix = {arXiv},
       eprint = {2510.23732},
 primaryClass = {astro-ph.HE},
       adsurl = {https://ui.adsabs.harvard.edu/abs/2025arXiv251023732K}
}

@ARTICLE{HallX+25_AT2025ulz_S250818k,
       author = {{Hall}, Xander J. and {Busmann}, Malte and {Koehn}, Hauke and {Kunnumkai}, Keerthi and {Palmese}, Antonella and {O'Connor}, Brendan and {Freeburn}, James and {Hu}, Lei and {Gruen}, Daniel and {Dietrich}, Tim and {Bulla}, Mattia and {Coughlin}, Michael W. and {Antier}, Sarah and {Pillas}, Marion and {Price}, Paul A. and {Ahumada}, Tom{\'a}s and {Amsellem}, Ariel and {Andreoni}, Igor and {Augustin}, Jule and {Cabrera}, Tom'as and {Deshpande}, Rasika and {Fab{\`a}-Moreno}, Jennifer and {Gassert}, Julius and {Karpov}, Sergey and {Kasliwal}, Mansi and {Maga{\~n}a Hernandez}, Ignacio and {Mandelbaum}, Rachel and {Fontinele Nunes}, Felipe and {Pang}, Peter T.~H. and {Sommer}, Julian and {Stein}, Robert and {Tabor}, Constantin and {Vega}, Pablo and {Wouters}, Thibeau and {Zuo}, Xiaoxiong},
        title = "{AT2025ulz and S250818k: Investigating early time observations of a subsolar mass gravitational-wave binary neutron star merger candidate}",
      journal = {arXiv e-prints},
         year = 2025,
        month = oct,
          eid = {arXiv:2510.24620},
        pages = {arXiv:2510.24620},
          doi = {10.48550/arXiv.2510.24620},
archivePrefix = {arXiv},
       eprint = {2510.24620},
 primaryClass = {astro-ph.HE},
       adsurl = {https://ui.adsabs.harvard.edu/abs/2025arXiv251024620H}
}

@ARTICLE{Metzger_Hui_Cantiello24,
       author = {{Metzger}, Brian D. and {Hui}, Lam and {Cantiello}, Matteo},
        title = "{Fragmentation in Gravitationally Unstable Collapsar Disks and Subsolar Neutron Star Mergers}",
      journal = {\apjl},
         year = 2024,
        month = aug,
       volume = {971},
       number = {2},
          eid = {L34},
        pages = {L34},
          doi = {10.3847/2041-8213/ad6990},
archivePrefix = {arXiv},
       eprint = {2407.07955},
 primaryClass = {astro-ph.HE},
       adsurl = {https://ui.adsabs.harvard.edu/abs/2024ApJ...971L..34M}
}

@ARTICLE{Chen_Metzger25,
       author = {{Chen}, Yi-Xian and {Metzger}, Brian D.},
        title = "{Gravitational Instability and Fragmentation in Collapsar Disks Supports the Formation of Subsolar Neutron Stars}",
      journal = {\apjl},
         year = 2025,
        month = sep,
       volume = {991},
       number = {1},
          eid = {L22},
        pages = {L22},
          doi = {10.3847/2041-8213/ae045d},
archivePrefix = {arXiv},
       eprint = {2508.17183},
 primaryClass = {astro-ph.HE},
       adsurl = {https://ui.adsabs.harvard.edu/abs/2025ApJ...991L..22C}
}

@ARTICLE{Piro_Pfhal07,
       author = {{Piro}, Anthony L. and {Pfahl}, Eric},
        title = "{Fragmentation of Collapsar Disks and the Production of Gravitational Waves}",
      journal = {\apj},
         year = 2007,
        month = apr,
       volume = {658},
       number = {2},
        pages = {1173-1176},
          doi = {10.1086/511672},
archivePrefix = {arXiv},
       eprint = {astro-ph/0610696},
 primaryClass = {astro-ph},
       adsurl = {https://ui.adsabs.harvard.edu/abs/2007ApJ...658.1173P}
}

@ARTICLE{BelczynskiK+02,
       author = {{Belczynski}, Krzysztof and {Kalogera}, Vassiliki and {Bulik}, Tomasz},
        title = "{A Comprehensive Study of Binary Compact Objects as Gravitational Wave Sources: Evolutionary Channels, Rates, and Physical Properties}",
      journal = {\apj},
         year = 2002,
        month = jun,
       volume = {572},
       number = {1},
        pages = {407-431},
          doi = {10.1086/340304},
archivePrefix = {arXiv},
       eprint = {astro-ph/0111452},
 primaryClass = {astro-ph},
       adsurl = {https://ui.adsabs.harvard.edu/abs/2002ApJ...572..407B}
}

@ARTICLE{BelczynskiK+16,
       author = {{Belczynski}, Krzysztof and {Repetto}, Serena and {Holz}, Daniel E. and {O'Shaughnessy}, Richard and {Bulik}, Tomasz and {Berti}, Emanuele and {Fryer}, Christopher and {Dominik}, Michal},
        title = "{Compact Binary Merger Rates: Comparison with LIGO/Virgo Upper Limits}",
      journal = {\apj},
         year = 2016,
        month = mar,
       volume = {819},
       number = {2},
          eid = {108},
        pages = {108},
          doi = {10.3847/0004-637X/819/2/108},
archivePrefix = {arXiv},
       eprint = {1510.04615},
 primaryClass = {astro-ph.HE},
       adsurl = {https://ui.adsabs.harvard.edu/abs/2016ApJ...819..108B}
}

@ARTICLE{DaviesM+05,
       author = {{Davies}, Melvyn B. and {Levan}, Andrew J. and {King}, Andrew R.},
        title = "{The ultimate outcome of black hole-neutron star mergers}",
      journal = {\mnras},
         year = 2005,
        month = jan,
       volume = {356},
       number = {1},
        pages = {54-58},
          doi = {10.1111/j.1365-2966.2004.08423.x},
archivePrefix = {arXiv},
       eprint = {astro-ph/0409681},
 primaryClass = {astro-ph},
       adsurl = {https://ui.adsabs.harvard.edu/abs/2005MNRAS.356...54D}
}

@ARTICLE{StegmannJ_KlenckiJ25,
       author = {{Stegmann}, Jakob and {Klencki}, Jakub},
        title = "{Orbital Eccentricity and Spin─Orbit Misalignment Are Evidence that Neutron Star─Black Hole Mergers Form through Triple Star Evolution}",
      journal = {\apjl},
         year = 2025,
        month = oct,
       volume = {991},
       number = {2},
          eid = {L54},
        pages = {L54},
          doi = {10.3847/2041-8213/ae055b},
archivePrefix = {arXiv},
       eprint = {2506.09121},
 primaryClass = {astro-ph.HE},
       adsurl = {https://ui.adsabs.harvard.edu/abs/2025ApJ...991L..54S}
}

@ARTICLE{Zenati+25,
       author = {{Zenati}, Yossef and {Rozner}, Mor and {Krolik}, Julian H. and {Most}, Elias R.},
        title = "{Mass Transfer in Eccentric Black Hole{\textendash}Neutron Star Mergers}",
      journal = {\apj},
         year = 2025,
        month = jan,
       volume = {978},
       number = {2},
          eid = {126},
        pages = {126},
          doi = {10.3847/1538-4357/ad9b87},
archivePrefix = {arXiv},
       eprint = {2410.05391},
 primaryClass = {astro-ph.HE},
       adsurl = {https://ui.adsabs.harvard.edu/abs/2025ApJ...978..126Z}
}

@ARTICLE{CoyneR+2016,
       author = {{Coyne}, Robert and {Corsi}, Alessandra and {Owen}, Benjamin J.},
        title = "{Cross-correlation method for intermediate-duration gravitational wave searches associated with gamma-ray bursts}",
      journal = {\prd},
         year = 2016,
        month = may,
       volume = {93},
       number = {10},
          eid = {104059},
        pages = {104059},
          doi = {10.1103/PhysRevD.93.104059},
archivePrefix = {arXiv},
       eprint = {1512.01301},
 primaryClass = {gr-qc},
       adsurl = {https://ui.adsabs.harvard.edu/abs/2016PhRvD..93j4059C}
}

@ARTICLE{WuJ+26,
       author = {{Wu}, Jiaxi and {Most}, Elias R. and {Vu}, Nils L. and {Deppe}, Nils and {Kidder}, Lawrence E. and {Nelli}, Kyle C. and {Throwe}, William},
        title = "{Eccentricity as a signature of hierarchical subsolar-mass mergers in collapsar disks}",
      journal = {arXiv e-prints},
         year = 2026,
        month = apr,
          eid = {arXiv:2604.26912},
        pages = {arXiv:2604.26912},
          doi = {10.48550/arXiv.2604.26912},
archivePrefix = {arXiv},
       eprint = {2604.26912},
 primaryClass = {astro-ph.HE},
       adsurl = {https://ui.adsabs.harvard.edu/abs/2026arXiv260426912W}
}

@ARTICLE{Barnes&Metzger22,
       author = {{Barnes}, Jennifer and {Metzger}, Brian D.},
        title = "{Signatures of r-process Enrichment in Supernovae from Collapsars}",
      journal = {\apjl},
         year = 2022,
        month = nov,
       volume = {939},
       number = {2},
          eid = {L29},
        pages = {L29},
          doi = {10.3847/2041-8213/ac9b41},
archivePrefix = {arXiv},
       eprint = {2205.10421},
 primaryClass = {astro-ph.HE},
       adsurl = {https://ui.adsabs.harvard.edu/abs/2022ApJ...939L..29B}
}

@ARTICLE{PassalevaN+25_NIR,
       author = {{Passaleva}, N. and {Durbak}, J. and {Guiffreda}, O. and {Troja}, E. and {Hamada}, R. and {Kutyrev}, A.~S. and {Suzuki}, D. and {Sumi}, T. and {Idei}, A. and {Buckley}, D. and {Cenko}, S.~B.},
        title = "{LIGO/Virgo/KAGRA S250818k: PRIME nIR observations of AT2025ulz}",
      journal = {GRB Coordinates Network},
         year = 2025,
        month = aug,
       volume = {41504},
        pages = {1},
       adsurl = {https://ui.adsabs.harvard.edu/abs/2025GCN.41504....1P}
}

\end{document}